\documentclass[twocolumns,dvipdfmx]{pasj01}

\begin{document}

\SetRunningHead{Tosaki et al.} {GMCs in starburst ring of NGC1068}
%\Received\\pagestyle{empty}

{}%{yyyy/mm/dd}
%\Accepted{}%{yyyy/mm/dd}
%\Published{}%{yyyy/mm/dd}

\title{A statistical study of giant molecular clouds traced by $^{13}$CO, C$^{18}$O, CS, and CH$_3$OH in the disk of NGC 1068 based on ALMA observations}

%%% begin:list of authors
% Do NOT capitalize all letters in "textsc".
%\author{Tomoka \textsc{Tosaki} %
%  \thanks{Example: Present Address is xxxxxxxxxx}}
%\affil{Joetsu University of Education}
%\email{tosaki@juen.ac.jp}

%%% Please use the following style in case that sorting by
%%% affilation is impossible.
%

\author{%
   Tomoka \textsc{Tosaki}\altaffilmark{1},
   Kotaro \textsc{Kohno}\altaffilmark{2,3}
   Nanase \textsc{Harada}\altaffilmark{4}, 
   Kunihiko \textsc{Tanaka}\altaffilmark{5}, 
   Fumi \textsc{Egusa}\altaffilmark{6},
   Takuma  \textsc{Izumi}\altaffilmark{2},
   Shuro \textsc{Takano}\altaffilmark{7}, 
   Taku \textsc{Nakajima}\altaffilmark{8},
   Akio \textsc{Taniguchi}\altaffilmark{2},  
   Yoichi \textsc{Tamura}\altaffilmark{2}}

 \altaffiltext{1}{Joetsu University of Education, Yamayashiki-machi, Joetsu, Niigata 943-8512, Japan}\email{tosaki@juen.ac.jp}
  \altaffiltext{2}{Institute of Astronomy, School of Science, The University of Tokyo, Osawa, Mitaka, Tokyo 181-0015, Japan}
  \altaffiltext{3}{Research Center for Early Universe, School of Science, The University of Tokyo, Hongo, Bunkyo, Tokyo 113-0033, Japan}
 \altaffiltext{4}{Academia Sinica Institute of Astronomy and Astrophysics, P.O. Box 23-141, Taipei 10617,
Taiwan}
 \altaffiltext{5}{Department of Physics, Faculty of Science and Technology, Keio University, 3-14-1 Hiyoshi, Yokohama, Kanagawa 223–8522, Japan}
 \altaffiltext{6}{NAOJ Chile Observatory, National Astronomical Observatory of Japan, 2-21-1 Osawa, Mitaka, Tokyo 181-8588, Japan}
 \altaffiltext{7}{Department of Physics, General Studies, College of Engineering, Nihon University, Tamuramachi, Koriyama, Fukushima 963-8642, Japan}
 \altaffiltext{8}{Institute for Space-Earth Environmental Research, Nagoya University, 
Furo-cho, Chikusa-ku, Nagoya, Aichi 464-8601, Japan}

%% `\KeyWords{}' always has to be placed before `\maketitle'.
\KeyWords{galaxies: individual (NGC1068) -- galaxies: ISM -- radio lines: galaxies} %Do NOT move this preamble from here!

\maketitle

\begin{abstract}
We present $1''.4$ (98 pc) resolution ALMA observations of $^{13}$CO($J$=1--0), C$^{18}$O($J$=1--0), CS($J$=2--1) and CH$_3$OH($J_K$=$2_K$--$1_K$) molecular rotational lines in the central $1'$ (4.2 kpc) diameter region of NGC 1068 
to study the physical and chemical properties of giant molecular clouds (GMCs) and to test whether these GMC-scale properties are linked to the larger-scale galactic environment. 
Using the derived $^{13}$CO cube, we have identified 187 high-significance ($> 8\sigma$) GMCs by employing the \verb|CLUMPFIND| algorithm. 
The molecular gas masses of GMCs $M_{\rm ^{13}CO}$, derived from the $^{13}$CO data, range from $1.8 \times 10^4 ~ M_\odot$ to $4.2 \times 10^7 ~M_\odot$. 
A mass function of GMCs in NGC 1068 has been obtained for the first time at $\sim$100 pc resolution. We find the slope of the mass function $\gamma = -1.25 \pm 0.07$ for a mass range of $M_{\rm ^{13}CO} \geq 10^5M_\odot$. This is shallower than the GMCs in the disk regions of the Milky Way, M51 and NGC 300. Further, we find that the high mass cut-off of the GMC mass function occurs at $M_{\rm ^{13}CO} \sim 6 \times 10^7 M_\odot$, which is an order of magnitude larger than that in the nuclear bar region of M51, indicating that the more massive clouds dominate the mass budget in NGC 1068. 
The observed C$^{18}$O($J$=1--0)/$^{13}$CO($J$=1--0) intensity ratios are found to be fairly uniform (0.27 $\pm$ 0.05) among the identified GMCs. In contrast, the CH$_3$OH($J_K$=$2_K$--$1_K$)/$^{13}$CO($J$=1--0) ratios exhibit striking spatial variation across the disk, with the smallest values around the bar-end ($<$ 0.03), and larger ratios along the spiral arms ($\sim$ 0.1--0.2). We find that GMCs with detectable methanol emission tend to have systematically larger velocity widths than those without methanol emission, suggesting that (relatively weak) shocks are responsible for the enhancement of the CH$_3$OH/$^{13}$CO ratios of GMCs in the disk of NGC 1068. 

\end{abstract}

\section{Introduction}

The study of the nature and evolution of interstellar medium (ISM) is one of the indispensable steps 
for understanding the physical processes that govern star formation in the universe, 
because stars are formed in and return to it.
Molecular medium, which is a dense and cold phase of ISM, is the very site and material of star formation.
The molecular medium in \textcolor{black}{the Milky Way} and galaxies show a wide range of physical scales, from sub-parsec scale to a few hundred parsec scale. 
They are referred to cores, clumps, giant molecular clouds (GMCs), and giant molecular associations (GMAs) 
(\cite{Scoville1987}; \cite{Rand1990}).
In \textcolor{black}{the Milky Way}, 90\% of molecular gas is in the form of GMCs with a size of 20 pc or larger and a typical mass of $\sim4 \times 10^5 ~M_\odot$, and
massive stars are formed as a cluster in such GMCs (\cite{Scoville1987}; \cite{Scoville2013}; \cite{Lada2003}).
The physical and chemical properties of GMCs are greatly influenced by massive stars, owing to their stellar winds and supernovae 
\textcolor{black}{in the final stage};
%%after their death
for example, the surrounding ISM are compressed by them and new stars are formed there. 
Hence, it is essential to study the physical and chemical properties of GMCs, 
because these key processes will be imprinted into them.

Classically, it is well-known that the molecular clouds follow scaling relations such as the size-line width relation, a.k.a. Larson's law (\cite{Larson1981}; \cite{Solomon1987}; \cite{Rosolowsky2008}; \cite{Bolatto2008}), 
and star formation laws, such as the Schmidt-Kennicutt law (\textcolor{black}{the S-K law hereafter,} \cite{Schmidt1959}; \cite{Kennicutt1998}; \cite{Kennicutt2012}).
These are known to be very universal\textcolor{black}{, even in a high-redshift starburst galaxy (\citet{Swinbank2015} for the Larson's law, and \citet{Hatsukade2015} for the S-K law in the lensed starburst galaxy SDP.81 at $z=3.1$)}. However, recent studies suggest that such universality 
breaks down at some spatial scales and/or environments. 
For instance, the S-K law is shown to be valid only at scales larger than that of the GMCs, i.e., the breakdown of the S-K law occurs at $\lesssim$ 80 pc in the GMCs of M33 (\cite{Onodera2010}). 
\textcolor{black}{The size-line width relations are found to be different in the inner a few 100 pc regions of the \textcolor{black}{Milky Way} and galaxies such as NGC 253 (e.g., \cite{Leroy2016} and references therein; see also \citet{Utomo2015} for the absence of such clear size-line width correlation for the GMCs in NGC 4526, an early-type galaxy).}
%%{\bf No clear size-line width correlation was found for the GMCs in NGC 4526, an early-type galaxy, arguing that the difference on local environments such as higher interstellar radiation field than in the Milky Way disk may account for this (\cite{Utomo2015}).
\textcolor{black}{\citet{Hughes2010} have reported some of GMC properties in the Large Magellanic Cloud (LMC), such as the CO surface brightness, can vary depending on the local environments including the stellar mass surface density.}
Based on \textcolor{black}{the PdBI Arcsecond Whirlpool Survey} (PAWS), 
an extensive imaging survey of GMCs in M51 (\cite{Schinnerer2013}), 
the environmental dependence of the GMC mass spectrum has also been reported (\cite{Colombo2014}). \textcolor{black}{All these recent progresses are} motivating us to conduct more wider surveys of GMCs with a sufficiently high angular resolution (e.g., \cite{Leroy2016}).

These previous studies of GMCs in galaxies are mainly based on $^{12}$CO observations. 
%%{\bf There are some recent studies on GMCs in galaxis based on $^{12}$CO (\cite{Hughes2010}; \cite{DonovanMeyer2012}, 2013;  \cite{Rebolledo2012}, 2015). }
The $^{12}$CO emission is a bright, well-established proxy of the whole molecular gas, which is crucial for a statistical study of extragalactic GMCs. It is primarily very optically thick and molecular gas masses are derived by using a CO-to-H$_2$ conversion factor. It has been suggested that the CO-to-H$_2$ conversion factors in galaxies can vary by a factor of 10 or more depending on the environments of molecular clouds (see \cite{Bolatto2013} for a recent review), which can therefore introduce environment-dependent uncertainties. 
The isotopologue of CO, $^{13}$CO, gives 
%%an optically thin faithful tracer of molecular gas mass, 
\textcolor{black}{another empirical measure of molecular gas mass},
and extensive studies have been made using single-dish telescopes (e.g., \cite{Aalto1995}; \cite{Tosaki2002}; \cite{Matsushita2010}; \cite{Tan2011}; \cite{Vila-Vilaro2015}). However, only limited measurements of $^{13}$CO have been made so far for high- resolution, interferometric studies of extragalactic GMCs/GMAs (e.g., \cite{Helfer1995}; \cite{Papadopoulos1996}; \cite{{Aalto1997}}; \cite{Kikumoto1998}; \cite{Matsushita1998}; \cite{Meier2004}; \cite{Aalto2010}; \cite{Schinnerer2010}; \cite{Pan2015}; \cite{Alatalo2016}; \cite{Watanabe2016}; \cite{Konig2016}). This is particularly true for ``statistical'' studies of physical properties GMCs in the disk regions using $^{13}$CO;   
for example, 25 GMCs were identified in M64 based on $3''.5$ or 75 pc resolution $^{13}$CO($J$=1--0) observations using the BIMA array, which shows a different size-line width relationship from Larson's law (\cite{Rosolowsky2005a}). \citet{Hirota2011} conducted $\sim$50 pc resolution $^{13}$CO($J$=1--0) imaging of the northern spiral arm segment of IC 342 with the Nobeyama Millimeter Array and revealed that dissipation of excess kinetic energy of GMCs is a required condition for the onset of massive star formation based on 26 $^{12}$CO-identified GMCs. 
A significant development of such studies is highly expected in the ALMA era. 

GMC-scale chemical properties of extragalactic molecular clouds are also expected to bring new insights to the study of activities in galaxies. 
In the Galactic star forming regions, chemical evolution and spatial variation are found at core and clump scales (i.e., a few parsec or smaller), and it has been suggested that these chemical properties are intimately connected to star formation and associated processes such as outflows and resultant shocks 
(\textcolor{black}{e.g., \cite{Bachiller1997}; \cite{Sakai2007}; \cite{Ohashi2014}}).  
However, it is yet unexplored whether any chemical variation and evolution of ISM at a GMC scale (approximately a few tens to hundreds of parsecs) exists in disk regions of galaxies; 
\citet{Watanabe2016} studied chemical properties of five GMAs traced by C$^{18}$O at $\sim$300 pc resolution using CARMA, but the sensitivity and angular resolution are both not yet sufficient to address the variation of chemical properties and its link to the environment. Most of the  high-resolution interferometric multi-molecule observations of galaxies have been mainly made in the nuclear regions of active galactic nuclei (AGNs) such as the circumnuclear disks (CNDs) of NGC 1068 (e.g., \cite{Krips2011};　\cite{GarciaBurillo2014}; \cite{Viti2014}; \cite{Takano2014}; \cite{Nakajima2015}) and NGC 1097 (\cite{Martin2015}), the nuclear regions of star-forming galaxies (\cite{Meier2005}; \cite{Meier2015}) and (ultra) luminous infrared galaxies (e.g., \cite{Martin2011}; \cite{Sakamoto2013,Sakamoto2014};  \cite{Saito2015}; \cite{Costagliola2015}; \cite{Tunnard2015}; \cite{Lindberg2016}).   

Here we present high-resolution ($1''.4$ or 98 pc at a distance of 14.4 Mpc, \cite{Tully1988}; \cite{Bland-Hawthorn1997}) and sensitive $^{13}$CO($J$=1--0), C$^{18}$O($J$=1--0), CS($J$=2--1) and CH$_3$OH($J_K$=$2_K$--$1_K$) imaging of the central 1 arcmin (4.2 kpc) diameter region of NGC 1068, one of the nearest Seyfert/starburst hybrid galaxies in the local universe. 
We selected this galaxy because we find 
%%tentative
evidence for striking spatial variation of molecular line distributions among molecular species including $^{13}$CO, C$^{18}$O, CS, HC$_3$N, CH$_3$CN, and CH$_3$OH across the starbusting spiral arms in our previous (cycle 0) ALMA observations (\cite{Takano2014}).

Among these molecular lines, we focus on $^{13}$CO, C$^{18}$O, CS, and CH$_3$OH lines, and new observations were made during the ALMA cycle 2 operation. The angular resolution and sensitivity have been improved significantly compared with pre-ALMA $^{13}$CO and C$^{18}$O observations (e.g., \cite{Helfer1995}; \cite{Papadopoulos1996}; \cite{Krips2011}) and the ALMA cycle-0 measurement (\cite{Takano2014}). In particular, the noise level is improved by a factor of 6--10 (1$\sigma$ = 1.1 -- 1.7 mJy beam$^{-1}$ for a velocity resolution of $\sim$19 km s$^{-1}$ in \citet{Takano2014}, whereas the new observations reach a noise level of 0.6 mJy beam$^{-1}$ for a velocity resolution of 1.5 km s$^{-1}$; see section 2 for details). This significant improvement of sensitivity allows us to conduct the first extensive statistical study of physical and chemical properties of GMCs including methanol in disk regions of galaxies.

Methanol is formed by hydrogenation of CO on the surface of dust grains in a low temperature condition ($\sim$10--15 K; e.g., \cite{Watanabe2003}), and comes into a gas phase due to relatively weak shocks (\cite{Flower2012}; \cite{Viti2011}; \cite{Bachiller1995}). Therefore, methanol can be regarded as a tracer of weak shocks in molecular medium. Methanol emission has been detected \textcolor{black}{using interferometers} in the central region of the star-forming galaxy IC 342 (\cite{Meier2005}), \textcolor{black}{Maffei 2 (\cite{Meier2012})}, the interacting galax{ies} VV 114 (\cite{Saito2015}, \textcolor{black}{\yearcite{Saito2016}}), \textcolor{black}{NGC 4038 (\cite{Ueda2016})} and a spiral arm in M51 (\cite{Watanabe2016}); no $\leq$100 pc resolution observations in disk regions of galaxies have been reported so far. 

With these improved ALMA data sets, our goals are (1) to study the physical and chemical properties of giant molecular clouds (GMCs) based on the $^{13}$CO-selected cloud catalog, and (2) to test whether these GMC-scale properties are linked to the larger-scale galactic environment. Note that we focus on the properties of GMCs in the spiral arms and inter-arm regions in this study; specifically, $5'' < r < 30''$ or 350 pc $< r <$ 2.1 kpc from the nucleus (see section 3.3). This is because our selection criteria of clumps using $^{13}$CO($J$=1--0) emission are not optimized for clouds in the CND. The impact of the AGN on the surrounding dense molecular medium, i.e., physical and chemical properties of GMCs in the CND of NGC 1068 will be reported in a forthcoming paper. 

The paper is organized as follows. The ALMA observations and data reduction procedures are described in section 2. After the data presentation and flux comparison with the previous single-dish measurements, in \textcolor{black}{section} 3 we outline the methodology of cloud identification, and then the derived GMC catalog is presented along with the derivation of physical quantities of GMCs. In \textcolor{black}{section} 4, we discuss the physical and chemical properties of GMCs and their possible dependence on environment. The outcomes of this study are summarized in \textcolor{black}{section} 5. \textcolor{black}{The data products of this study, including the FITS file of the $^{13}$CO data cube, will be publicly available from a dedicated web page\footnote{http://www.juen.ac.jp/lab/tosaki/n1068/}.}

\section{Observations and Data Reduction}

The central $\sim$1 arcmin diameter region of NGC 1068 was observed 
using the Band 3 receiver on ALMA with the 2SB dual-polarization setup, 
as a cycle 2 early science program (Project code = 2013.0.00060.S; PI name = T.~Tosaki). 

The central frequencies of four spectral windows, i.e., spw0, spw1, spw2, and spw3, were tuned to 110.000 GHz, 109.174 GHz, 97.174 GHz, and 96.741 GHz, respectively. 
Each spectral window has a bandwidth of 1.875 GHz with 3840 channels, giving a total frequency coverage of 7.5 GHz and a spectral resolution of 0.488 MHz. 
This configuration allows us to simultaneously observe the $J$=1--0 lines of $^{13}$CO (the adopted rest frequency $f_{\rm rest}$ = 110.201353 GHz, \cite{Lovas1992}) and C$^{18}$O (109.782160 GHz) in the upper sideband (USB), 
and the $J$=2--1 line of CS (97.980968 GHz) and $J_K$=$2_K$--$1_K$ group lines of CH$_3$OH (96.74142 GHz)\footnote{Transitions contributing to this $2_K$--$1_K$ group are $J_{K_a,K_c}$=$2_{-1,2}$--$1_{-1,1}E$, $2_{0,2}$--$1_{0,1}A$+, $2_{0,2}$--$1_{0,1}E$, and $2_{1,1}$--$1_{1,0}E$. The indicated rest frequency in the text is for $2_{0,2}$--$1_{0,1}A$+.} in the lower sideband (LSB), 
along with other rarer species such as HNCO and HC$_3$N (not presented in this paper). 

The ALMA observations were taken during three separated periods with six execution blocks in total. Table \ref{table:1} summarizes the details of the ALMA observations. 
Bright quasars including J0108+0135, J0224+0659, J0238+1636, J0241-0815, and J0423-0120 were observed to calibrate pass-band characteristics, and 
Mars, Uranus, and Neptune were used as primary flux calibrators. The resultant accuracy of the absolute flux scale is reported to be better than 10\% according to the ALMA Technical Handbook\footnote{\textcolor{black}{https://almascience.nao.ac.jp/documents-and-tools/cycle4/alma-technical-handbook}}.

Calibration of the raw visibility data was conducted using the Common Astronomy Software Applications (CASA; \cite{McMullin2007}; \cite{Petry2012}). Version 4.3.1  
\textcolor{black}{of the} program was used for the first two data sets, whereas the remaining four data sets were processed using the version 4.2.2 with pipeline. 

The image reconstruction from the calibrated visibilities and deconvolution were done with the task \verb|CLEAN| in CASA version 4.5.0. After the initial imaging with natural weighting, we determined the line-free channels and then continuum emission was subtracted from the visibilities using the task \verb|uvcontsub|. 
In order to obtain a similar beam size for the target lines with different frequencies, i.e., $^{13}$CO($J$=1--0) and C$^{18}$O($J$=1--0) in the USB, and CH$_3$OH($J_K$=$2_K$--$1_K$) and CS($J$=2--1) in the LSB, we created images using Briggs weighting with different robust parameters: \verb|robust| = 0.4 for $^{13}$CO and C$^{18}$O, 0.2 for CS, and 0.1 for CH$_3$OH. These parameters give the native beam sizes of $1''.39 \times 1''.09$, $1''.33 \times 1''.06$, $1''.38 \times 1''.12$, and $1''.39 \times 1''.12$ for $^{13}$CO, C$^{18}$O, CS, and CH$_3$OH, respectively. The CLEANed images were then finally convolved to a circular Gaussian with the FWHM of $1''.4$, giving a spatial resolution of 98 pc at the adopted distance of NGC 1068 (14.4 Mpc). 

Regarding the deconvolution, we employed the multi-scale CLEAN algorithm; the conventional CLEAN algorithm models an image by a sum of point sources (i.e., Gaussians with a single FWHM value), whereas the multi-scale CLEAN algorithm models an image by a collection of multiple Gaussians with different spatial scales. This method is known to recover higher fraction of flux for extended emission than the conventional CLEAN algorithm (\cite{Cornwell2008}; \cite{Rich2008}). We adopt the \verb|multiscale = [0, 10, 25]| with \verb|cell = '0.25arcsec'| in the task \verb|CLEAN|, specifying spatial scales of the synthesized beam size (i.e., the scale used in the conventional CLEAN), $2''.5$, and $6''.25$. This setup roughly corresponds to the scales of a point source (i.e., the beam size), the beam size $\times$2, and the beam size $\times$5, which are recommended in the CASA guide\footnote{https://casa.nrao.edu/Release4.1.0/doc/UserMan/UserMansu270.html}.

The resultant 1$\sigma$ noise levels in the channel maps, which are measured in the line-free channels, are 0.64, 0.61, 0.59, and 0.70 mJy beam$^{-1}$ for $^{13}$CO, C$^{18}$O, CS, and CH$_3$OH, respectively, at a velocity resolution of 1.5 km s$^{-1}$.

\section{Results}

\subsection{Global distributions of molecular \textcolor{black}{emission lines}}

\begin{figure*}[H]
 \begin{center}
    \FigureFile(180mm,180mm){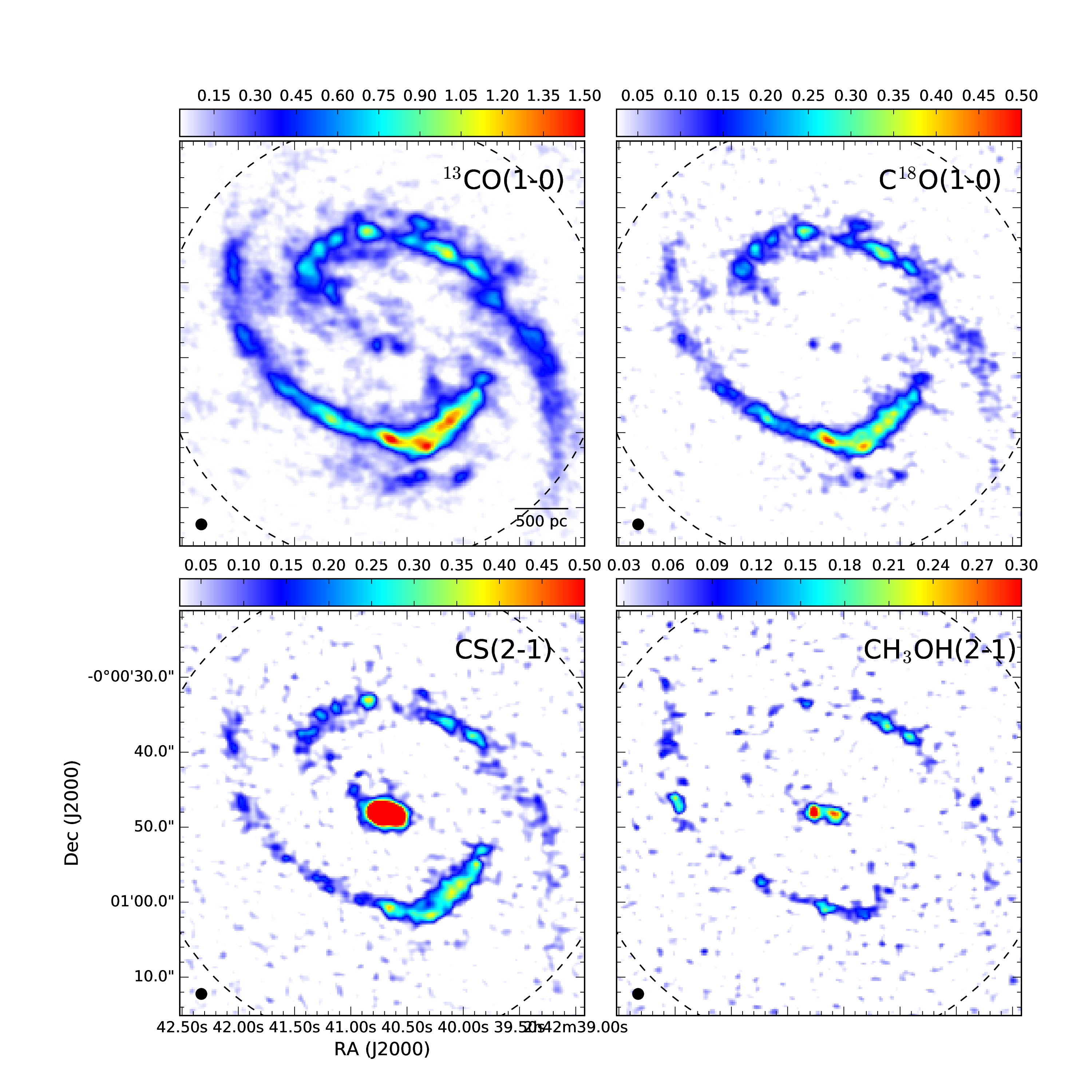}
    %%% \FigureFile(width,height){filename}
  \end{center}
\caption{Velocity-integrated intensity images of $^{13}$CO($J$=1--0), C$^{18}$O($J$=1--0), CS($J$=2--1), and CH$_3$OH($J_K$=$2_K$--$1_K$), obtained with the Cycle-2 ALMA. These images have a common synthesized beam size, $1''.4$ (circular), as shown in the bottom left corner of each panel. Dashed circles show the field of view. Attenuation of ALMA 12-m telescope primary beam pattern is not corrected in these figures. Color bars of each panel show the velocity-integrated intensity in a unit of Jy beam$^{-1}$ km s$^{-1}$. The typical noise level of these moment-0th images is $\sim$0.015 Jy beam$^{-1}$ km s$^{-1}$.  
}\label{fig:1}
\end{figure*}

The total integrated intensity maps of $^{13}$CO($J$=1--0), C$^{18}$O($J$=1--0), CS($J$=2--1) and CH$_3$OH($J_K$=$2_K$--$1_K$) emission lines are displayed in figure \ref{fig:1}. We computed 0th moment images of these molecular lines by using the \verb|CASA| task \verb|immoments|, over a velocity range of $V_{\rm LSR}$ = 955--1315 km s$^{-1}$ without any clipping to obtain these images. 

We detect all four emission lines in the CND and the starburst ring or tightly winding two-armed spirals. 
In particular, the high dynamic range ($\sim$72 in the channel maps) $^{13}$CO image, 
which is produced with a wide range of baseline length (see table \ref{table:1}, corresponding to 5.5 k$\lambda$ -- 290 k$\lambda$), excellent uv-coverage accomplished with multiple configuration observations with 30--40 antennas, and the multi-scale CLEAN implemented in \verb|CASA|, successfully reveals detailed sub-structures of spiral arms and faint clouds in the inter-arm regions, which are similar to recent fully-sampled sensitive $^{12}$CO observations of M51 (\cite{Koda2009}; \cite{Schinnerer2013}) and often reproduced in state-of-the-art numerical simulations (e.g., \cite{Baba2013}). 

The \textcolor{black}{emission lines} of $^{13}$CO and C$^{18}$O in the CND are relatively weak, whereas we find striking peaks in CS and CH$_3$OH, as pointed out by the previous ALMA cycle 0 observations (\cite{Takano2014}). 
The positions of peak intensities in the $^{13}$CO, C$^{18}$O, and CS maps are similar to each other, whereas the distribution of prominent CH$_3$OH peaks is significantly different from these three lines. This point will be discussed more quantitatively in section \ref{sec:GMC_properties}.

\subsection{Comparison of flux with single-dish telescopes}

The $^{13}$CO($J$=1--0) flux obtained from ALMA were compared with those from single-dish telescopes (NRO 45-m and IRAM 30-m) to estimate the recovery fraction of the total line flux. 

First, the $^{13}$CO($J$=1--0) data cube was convolved to $16''$, the beam size of the NRO 45-m telescope at 110 GHz.
We then obtained a spectrum at the position of the nucleus, which was the pointing position of the 45-m telescope observation, and the velocity-integrated flux was computed. The resultant $^{13}$CO($J$=1--0) flux is 18.8 Jy km s$^{-1}$. On the other hand, a NRO 45-m telescope measurement gives a $^{13}$CO($J$=1--0) flux of $8.9\pm0.2$ K km s$^{-1}$ in the main-beam temperature scale (Takano et al., in preparation), corresponding to a flux density of $22.5\pm 0.51$ Jy km s$^{-1}$. If we consider the absolute flux calibration accuracy of the NRO 45-m telescope (typically $\pm 15$~\%, e.g., \cite{Yoshida2015}) and ALMA ($\pm 5$\%, the ALMA Technical Handbook\footnote{\textcolor{black}{https://almascience.nao.ac.jp/documents-and-tools/cycle4/alma-technical-handbook}}), the ratio of these two flux measurements is $\textcolor{black}{(18.8 \pm 0.9)}/(22.5 \pm 3.4) = 0.84 \pm 0.13$.

Similarly, we then compare the ALMA flux with that using the IRAM 30-m telescope in the same manner; the ALMA $^{13}$CO cube was convolved to $21''$, the 30-m beam size at 110 GHz, and we obtained a spectrum at the center to compute the integrated intensity. It was found to be $37.2\pm1.86$ Jy km s$^{-1}$, whereas the IRAM 30-m measurement gave $13.10 \pm 0.24$ K km s$^{-1}$ (\cite{Aladro2015}) or $57.0 \pm 1.0$ Jy km s$^{-1}$. Again, considering a typical absolute calibration accuracy of IRAM 30-m telescope ($\sim \pm10$\%, EMIR Users Guide\footnote{http://www.iram.es/IRAMES/mainWiki/EmirforAstronomers}), the fraction of the recovered flux is then $\textcolor{black}{(37.2 \pm 1.9)/(57.0 \pm 5.7) = 0.65 \pm 0.07}$. 

These two measurements of the recovered flux ratios, i.e., ALMA/45-m = $84 \pm 13$\%, and ALMA/30-m = \textcolor{black}{$65 \pm 7$\%}, suggest that the overall $^{13}$CO flux seems to be recovered in these ALMA observations, despite the lack of Atacama Compact Array (ACA, a.k.a.~Morita Array) measurements. Note that almost all single-dish flux is claimed to be recovered in the ALMA cycle 0 observations of CO($J$=3--2) in NGC 1068 without ACA measurements (\cite{GarciaBurillo2014}).

Although the difference between the recovered flux ratios of ALMA/45-m and ALMA/30-m is 
%%marginal
\textcolor{black}{not significant}, 
such a difference can happen indeed, if we recall the spatial extent of the $^{13}$CO($J$=1--0) emission in the central region of NGC 1068 
and the difference of the NRO 45-m and IRAM 30-m beam sizes; 
a significant fraction of $^{13}$CO($J$=1--0) emission comes from the 
%%circumnuclear 
starburst ring with a diameter of $\sim 30''$, and the IRAM 30-m beam ($21''$) can couple with the emission from the starburst ring. Such extended emission tends to be resolved out 
by interferometric observations, giving a smaller recovered flux for IRAM measurements. 
On the other hand, the beam size of NRO 45m observations ($16''$) is significantly smaller 
than the size of the 
%%circumnuclear 
starburst ring, and it mainly catches the $^{13}$CO($J$=1--0) emission 
in the CND, which is spatially compact ($<$ $4''$), resulting in a higher recovered flux ratio. 
By adding ACA (including total power array) measurements to our ALMA data in the near future, we will be able to obtain fully sampled $^{13}$CO (and other molecular lines) data sets to confirm such estimations quantitatively. 

In any case, we suggest that the missing flux (if any) will have rather minor impact on the following analysis, because we mainly focus on the clumpy structures, i.e., GMCs, in this region.

\subsection{Cloud identification using $^{13}$CO data cube}

We applied the \verb|CLUMPFIND| algorithm (\cite{Williams1994}; \cite{Rosolowsky2005a}) to the $^{13}$CO($J$=1--0) 
data cube in order to decompose the $^{13}$CO emission into discrete clouds and measure their physical properties. 
We adopt a peak threshold of 8$\sigma$ (5.12 mJy beam$^{-1}$) and 
a contour increment of 4$\sigma$. 
This resulted in the detection of 265 clumps. After excluding clumps with the derived 
clump diameter smaller than $1''.4$ (i.e., the CLEAN beam), 
we constructed a catalog of clouds containing 187 individual clumps in the observed field
of NGC 1068. 

In this study, we adopted a fairly high cut-off level (i.e., a threshold of 8$\sigma$ and a contour increment of 4$\sigma$), for the following reasons. 
Firstly, one of our goals of this study is to measure much weaker \textcolor{black}{emission lines} from rarer species such as C$^{18}$O and methanol, so the base GMC sample using $^{13}$CO should be detected in high signal-to-noise ratios (S/N). Secondly, another goal is to study the GMC mass spectrum in this region, and in order to compare with the previous studies, the molecular gas mass range of our prime interest is $\sim 10^4$ $M_\odot$ or larger (see section \ref{section:GMC_Mass_Function}); the adopted threshold of 8$\sigma$ corresponds to a molecular gas mass of $\sim 1.7 \times 10^4~ M_\odot$ (for a cloud with a size comparable to the beam size), and therefore the high cut-off is sufficient for this purpose, too. Lastly, the \verb|CLUMPFIND| algorithm was originally implemented for high S/N data, 
and some issues are known when applying it to the data with a low dynamic range 
(\cite{Rosolowsky2005a}; \cite{Sheth2008}). By setting such a high threshold, we can construct a robust GMC catalog which is less affected by software issues. 
It should also be emphasized that high velocity resolution (1.5 km s$^{-1}$) 
is essential for the study of molecular gas in the disk region of galaxies
because the line widths of molecular clouds in the galactic disks can 
be just a few km s$^{-1}$, as shown in the later sections. 
We also note that because of the high S/N of GMCs, the accuracy of positions of the identified GMCs shall be significantly better than the beam size $\theta_{\rm beam}$ (as it scales as $\propto \theta_{\rm beam}$/(S/N)). 

The choice of the analysis algorithms, i.e., CPROPS (\cite{RosolowskyLeroy2006}) and dendrogram (\cite{Rosolowsky2008}), can be another issue \textcolor{black}{(e.g., \cite{DonovanMeyer2013})}. Intensity-based approaches, rather than decomposing the emission into clumps, have also been proposed (\cite{Sawada2012}; \cite{Hughes2013}; \cite{Leroy2016}). Detailed analysis of clump identification and/or intensity-based characterization of $^{13}$CO-identified cloud properties down to the sensitivity limit (e.g., 5$\sigma$ or less) with multiple-algorithm and approach comparison is beyond the scope of this paper, and such analysis will be presented elsewhere.

\subsection{Physical properties and molecular line ratios of the identified clouds}
\label{sec:M_13CO-Mvir}

We estimate the molecular gas mass of each cloud from the velocity-integrated $^{13}$CO intensity, $M_{\rm ^{13}CO}$, using the following equation, which assumes local thermodynamic equilibrium (LTE) and optically thin emission (e.g., \cite{Meier2000}), 
\begin{eqnarray}
%\begin{split}
N({\rm H_2})_{^{13}{\rm CO}}({\rm cm}^{-2}) = 2.41 \times 10^{14} \frac{[{\rm H}_2]}{[^{13}{\rm CO}]} \frac{e^{5.29/T_{\rm ex}(K)}}{e^{5.29/T_{\rm ex}(K)}-1} & \nonumber\\
\times I_{^{13}{\rm CO}}  ({\rm K~ km~s}^{-1}), 
%\end{split}
\end{eqnarray}
where $N({\rm H_2})$ is the molecular hydrogen column density. Adopting an  
abundance ratio of $[{\rm H}_2]/[^{13}{\rm CO}] = 5.0 \times 10^5$ (\cite{Dickman1978}) 
and an excitation temperature $T_{\rm ex}$ of 8.5 K (\cite{Nakajima2015}), 
we obtain the column density, and it is then multiplied by the cloud area, 
\textcolor{black}{which is the projected area of the clump on the sky obtained by the} \verb|CLUMPFIND|,   
to obtain the molecular gas mass $M_{^{\rm 13}{\rm CO}}$. 
We obtain the molecular gas masses of GMCs $M_{^{\rm 13}{\rm CO}}$ in NGC 1068 ranging from $1.8 \times 10^4$ to $4.2 \times 10^7 M_\odot$. 
%%ranging from $1.82 \times 10^4$ to $4.21 \times 10^7 M_\odot$. 
%%The averaged $M_{^{\rm 13}{\rm CO}}$ is $(2.87 \pm 5.38)\times 10^6 M_\odot$. 
The averaged $M_{^{\rm 13}{\rm CO}}$ is 2.9 $\times 10^6 M_\odot$. 

Assuming a spherical cloud with a density profile $\rho \propto r^{-\beta}$, we calculate the virial mass of GMC using the following formula (e.g., \cite{MacLaren1988}\textcolor{black}{; see also \cite{Hirota2011}; \cite{DonovanMeyer2012}}), 
\begin{eqnarray}
M_{\rm vir} (M_\odot) = 126 \frac{5-2\beta}{3-\beta} \left(\frac{R}{{\rm pc}}\right) \left(\frac{\Delta v}{{\rm km~s}^{-1}}\right)^2, 
\end{eqnarray}
where $R$ is the radius of the cloud \textcolor{black}{(the effective circular radius defined by \citet{Williams1994}, which is deconvolved by the beam size)} and $\Delta v$ is the velocity width of the $^{13}$CO profile of the cloud. Here we adopt $\beta=1$ for consistency with previous similar measurements. The derived virial masses have a range from $2.7\times 10^4$ to $1.7\times 10^7 M_\odot$, and the average value, 
%%virial masses have a range from $2.66\times 10^4$ to $1.69\times 10^7 M_\odot$, and the average value, 
%%$(2.59\pm 2.85) \times 10^6 M_\odot$, 
$2.6 \times 10^6 M_\odot$,
is similar to that of $M_{^{13}{\rm CO}}$.

The derived physical parameters of GMCs, i.e., position in the 3D-cube (dR.A., dDecl., and velocity), peak flux of $^{13}$CO emission, cloud radius, and velocity width (the full width of half maximum or FWHM of the $^{13}$CO($J$=1--0) spectrum) of 187 GMCs are listed 
in table \ref{table:2}, 
along with the molecular gas mass estimated from the $^{13}$CO emission and virial mass.
Note that the cloud position, dR.A. and dDecl., means the distance from the center of Field of View, (R.A.(J2000) = \timeform{2h42m40.7s}, Decl.(J2000) = \timeform{-0D0'47.9''}).

\begin{figure*}[H]
  \begin{center}
    \FigureFile(150mm,150mm){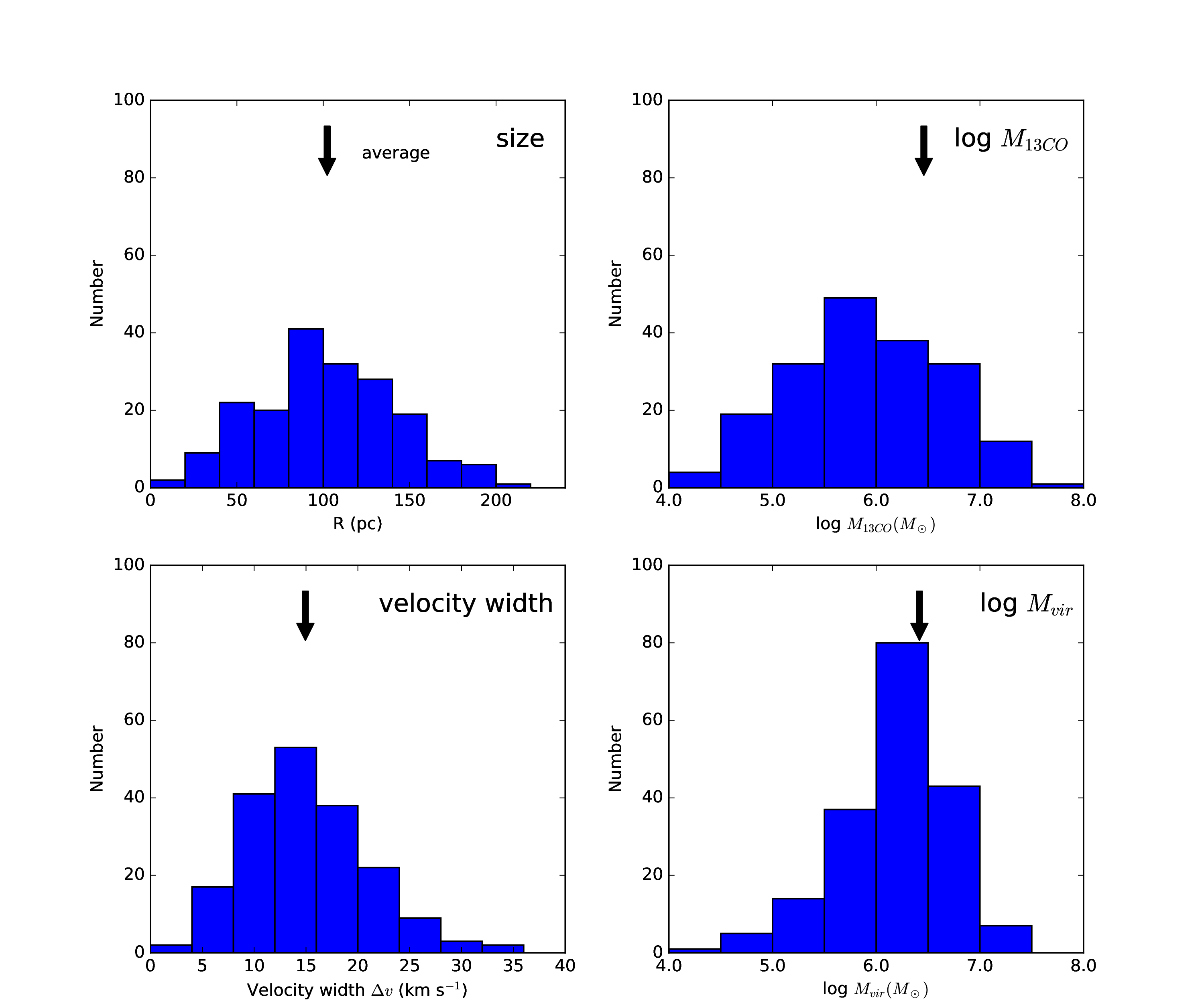}
    %%% \FigureFile(width,height){filename}
  \end{center}
\caption{Histograms of the sizes (top left), velocity widths (bottom left), molecular gas mass ($M_{\rm ^{13}CO}$, top right) and virial mass ($M_{\rm vir}$, bottom right) of the identified clouds. The average value of each quantity is indicated by an arrow. }\label{fig:2}
\end{figure*}

Figure \ref{fig:2} shows histograms of the size, velocity width, molecular gas mass ($M_{^{13}{\rm CO}}$), and virial mass for the identified clouds. 
%Table \ref{table:4} summarizes averaged values and ranges of these physical quantities for the identified clouds. 
The averaged size and velocity width of the identified clouds are 102 pc and 14.9 km s$^{-1}$, respectively. 
These value are slightly larger than the typical value for GMCs in the Milky Way (\cite{Sanders1985}), 
and smaller than those of GMAs in other galaxies (e.g. \cite{Rand1990}, \cite{Muraoka2009}). 
We refer to the identified clouds as GMCs. 

\begin{figure*}[H]
  \begin{center}
    \FigureFile(150mm,150mm){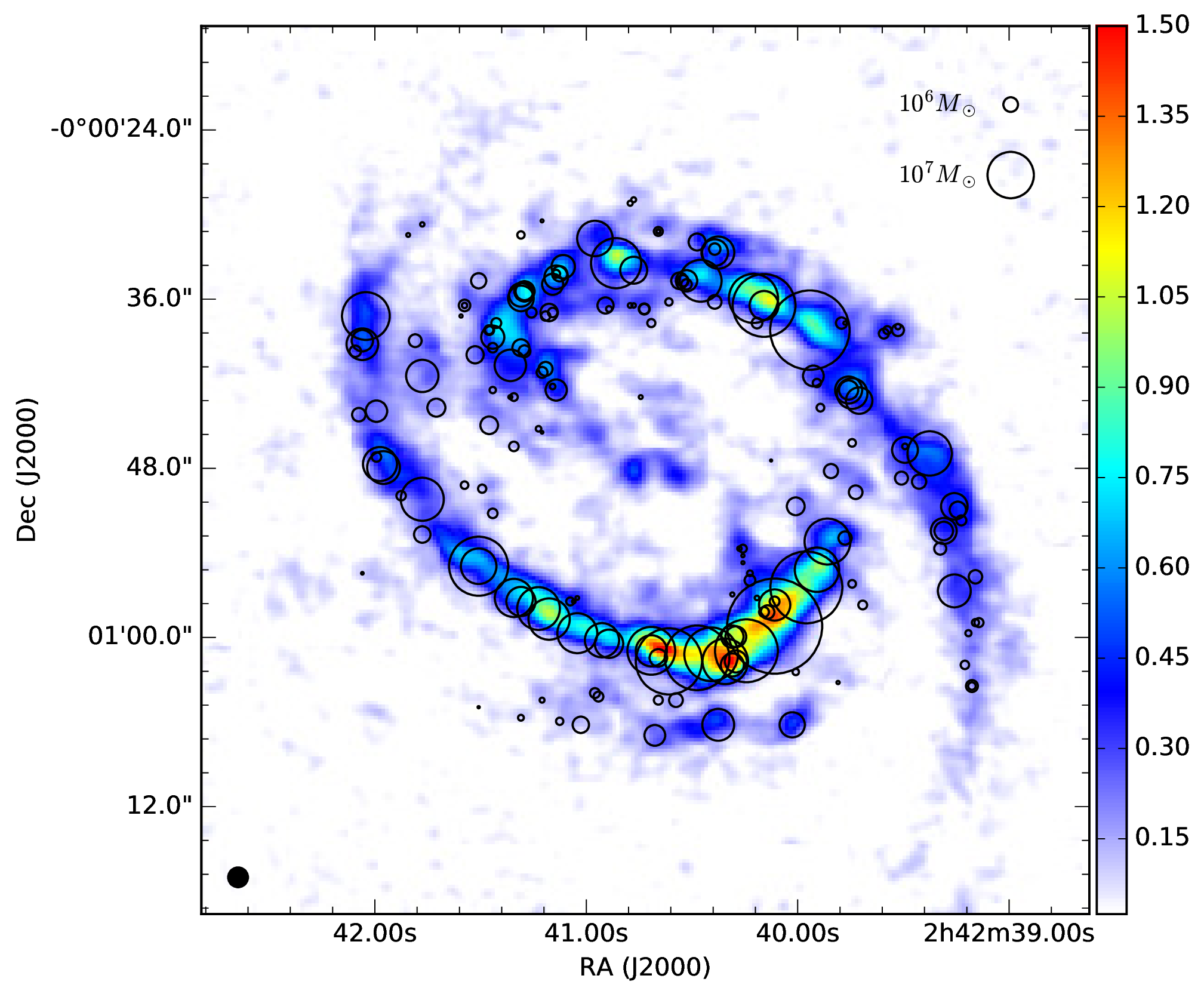}
    %%% \FigureFile(width,height){filename}
  \end{center}
  \caption{Positions of 187 identified clouds superposed on the $^{13}$CO($J$=1--0) integrated intensity image (the same as the top-left panel of figure 1). The color bar in the right shows the integrated intensity scale in a unit of Jy beam$^{-1}$ km s$^{-1}$. The 1$\sigma$ noise level of this map is 0.015 Jy beam$^{-1}$ km s$^{-1}$, or 0.78 K km s$^{-1}$, and the peak value of the map is 1.58 Jy beam$^{-1}$ km s$^{-1}$ or 81.7 K km s$^{-1}$. The synthesized beam size, $1''.4$ (circular) or 98 pc, is shown at the bottom-left corner. The size of circles is proportional to the molecular gas mass (LTE masses) of GMCs derived from the $^{13}$CO($J$=1--0) intensity. The molecular gas masses of the identified GMCs $M_{\rm ^{13}CO}$, derived from the $^{13}$CO data, range from $1.8 \times 10^4 ~ M_\odot$ to $4.2 \times 10^7 ~M_\odot$. 
Two circles, corresponding to the molecular gas masses of $10^6 ~ M_\odot$ and $10^7 ~ M_\odot$, respectively, are displayed at the top-right corner for reference. 
Note that no clumps are identified in the circumnuclear disk (CND) region ($r<5''$ or 350 pc) in the applied criteria (8$\sigma$ threshold and 4$\sigma$ increment), although significant ($>5\sigma$) emission is detected there. }\label{fig:3}
\end{figure*}

\begin{figure*}[H]
  \begin{center}
    \FigureFile(180mm,180mm){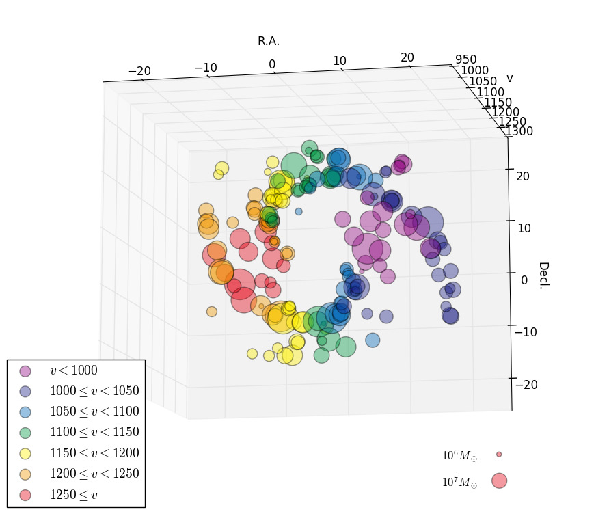}
    %%% \FigureFile(width,height){filename}
  \end{center}
  \caption{A 3-dimensional view of the identified cloud positions. The size of circles is proportional to the molecular gas mass (LTE mass) of the identified GMCs. Different colors of circles indicate the cloud peak velocities ($V_{\rm LSR}$). These clouds are often overlapped in figure 3, but separated by velocity as shown here. }\label{fig:4}
\end{figure*}

Here we display the identified clouds on the $^{13}$CO integrated intensity map in figure \ref{fig:3}. The identified clouds are located in the $5'' < r < 30''$ or 350 pc $< r <$ 2.1 kpc region from the nucleus. 
Note that no GMCs were identified in the central region including the CND \textcolor{black}{under 
%%the 
our criterion}, although we do detect significant ($>5\sigma$) $^{13}$CO emission in the CND. The physical properties of GMCs in the CND, which are expected to be impacted by activities from the central AGN, will be reported in a separate paper. Although a number of clouds are overlapped in this 2-dimensional view, figure \ref{fig:4} gives a clearer 3-dimensional view of the positions of the identified clouds. 
In figures \ref{fig:3} and \ref{fig:4}, the sizes of the symbols are proportional to their molecular gas mass $M_{^{13}{\rm CO}}$. 
Although most GMCs are located on the starburst ring, some GMCs are also seen outside of the ring. We find a wide range of GMC masses on the starburst ring, whereas GMCs outside the ring are predominantly less massive. 

Table \ref{table:3} shows the intensities of $^{13}$CO($J$=1--0), C$^{18}$O($J$=1--0), CS($J$=2--1), CH$_3$OH($J_K$=$2_K$--$1_K$) and 
the virial parameters for each GMC.
These molecular line intensities in table \ref{table:3}  
are derived from the integrated intensity over pixels of each GMC.
187, 182, and 121 GMCs with peak intensities of larger than $3\sigma$ are detected for C$^{18}$O, CS and CH$_3$OH, respectively.
Figure \ref{fig:5} shows histograms of the line intensity ratios for the GMCs.
The histograms of integrated intensity ratios show that C$^{18}$O/$^{13}$CO and CS/$^{13}$CO have a single peaked distribution. 
%%It is very clearly shown that the diameters of circles representing C$^{18}$O/$^{13}$CO ratios 
%%are almost uniform across the starburst ring, as expected from the narrow variation of the ratio. 
The CS/$^{13}$CO ratios show a similar trend to the C$^{18}$O/$^{13}$CO ratios, although their scatter is larger than the C$^{18}$O/$^{13}$CO ratios. 
On the other hand, CH$_3$OH/$^{13}$CO shows a wide-spread distribution ($\leq$  0.01 to 0.1--0.2). 
Table \ref{table:4} shows the averaged values  and ranges of intensit\textcolor{black}{y} ratios and virial parameters for these clouds together with their physical properties.
 
Here we mention the possible systematic uncertainties of these physical quantities. The molecular gas mass estimated from $^{13}$CO, $M_{\rm ^{13}CO}$, can have two possible error sources: the $^{13}$CO fractional abundance, $[{\rm H}_2]/[^{13}{\rm CO}]$, and the adopted excitation temperature $T_{\rm ex}$. The adopted $^{13}$CO fractional abundance (\cite{Dickman1978}) is based on the measurements of GMCs in the Galactic disk regions, and hence it is not trivial to determine if such a value is applicable to the starburst ring of NGC 1068. Future direct estimation of the $^{13}$CO fractional abundance will be necessary. Regarding the excitation temperature of $^{13}$CO, it is thought to be reliable because this is based on the measurements of 
%%multi-$J$
$^{13}$CO \textcolor{black}{$J=$3--2 and 1--0 lines} in the starburst ring (\cite{Nakajima2015}). Nevertheless, some cloud-to-cloud variation of $T_{\rm ex}$ may exist, depending on the star-forming activities of each cloud. For instance, some spatial variation of the CO($J$=3--2)/CO($J$=1--0) ratios has been reported across the starburst ring of NGC 1068 (\cite{GarciaBurillo2014}), implying a possible variation of physical parameters. Again, we may need additional observations such as $J$=2--1 and/or 3--2 of $^{13}$CO with a similar angular resolution, to make accurate measurements of $T_{\rm ex}$ of individual clouds in the future.

\section{Discussion}

\subsection{Comparison with GMCs in the Milky Way and local galaxies}

The size - velocity width relation of identified GMCs is shown in figure \ref{fig:6}, which also shows the correlation of molecular clouds in the Milky Way (\cite{Sanders1985}, based-on $^{12}$CO($J$=1--0)), cores/clumps in the Galactic star-forming region W51 (\cite{Parsons2012}, $^{13}$CO($J$=3--2)), and GMCs/GMAs in local galaxies including LMC (\cite{Fukui2008}, $^{12}$CO($J$=1--0)), M33 (\cite{Onodera2012}, $^{12}$CO($J$=1--0)), M51 (\cite{Colombo2014}, $^{12}$CO($J$=1--0)) and M83 (\cite{Muraoka2009}, $^{12}$CO($J$=3--2)). \textcolor{black}{This plot demonstrates that the Larson's relation observed with $^{13}$CO in the central region of NGC 1068 exhibits overall agreement with those measured with $^{12}$CO in local spiral galaxies (e.g., \cite{Gratier2012}; \cite{DonovanMeyer2013}; \cite{Rebolledo2015}).}
We found a trend for the GMCs with large velocity width in NGC 1068 similar to GMCs in the Milky Way, though the scatter is large.
One of the possible cause of this scatter could be an effect of size overestimation for small GMCs due to the spatial resolution in the \verb|CLUMPFIND| algorithm.
\textcolor{black}{We also note that $^{12}$CO emission tends to trace the outskirts of GMCs better than $^{13}$CO, so the extents of
 GMCs defined in $^{12}$CO and $^{13}$CO could be different. A high cut-off level of cloud identification in $^{13}$CO 
 may also give a smaller cloud size than $^{12}$CO measurements, although these effects are not clearly evident 
 in this figure.}

Figure \ref{fig:7} shows the correlation between LTE mass ($M_{\rm ^{13}CO}$) and virial mass of the identified GMCs. 
%%The correlation of M33, M51 and M83 in figure \ref{fig:7} are obtained by 
%%$M_{{\rm ^{12}CO}(J=1-0)}$, $M_{{\rm ^{12}CO}(J=2-1)}$, and $M_{{\rm ^{12}CO}(J=3-2)}$,  respectively (\cite{Onodera2012}; \cite{Colombo2014}; \cite{Muraoka2009}).
The virial masses in NGC 1068 are proportional to $M_{\rm ^{13}CO}$ as well as those of GMCs 
in the Milky Way \textcolor{black}{(W51)}, LMC, M33, and M83, although $M_{\rm vir}$ is slightly larger than $M_{^{13}{\rm CO}}$.
This suggests that the most of the \textcolor{black}{observed} GMCs 
%in this region
\textcolor{black}{, traced by $^{13}$CO,} are self-gravitating\textcolor{black}{, as found in recent high-resolution interferometric $^{12}$CO observations of GMCs in local galaxies (e.g., \cite{DonovanMeyer2012}, 2013; \cite{Colombo2014}; \cite{Rebolledo2012}, 2015; \cite{Utomo2015}; \cite{Faesi2016})}.

\subsection{Mass function of GMCs in the disk of NGC 1068}
\label{section:GMC_Mass_Function}

The mass spectrum or mass distribution of GMCs is one of the fundamental characteristics of molecular medium in \textcolor{black}{the Milky Way} and galaxies (e.g., \cite{Rosolowsky2005b}). The GMC mass distribution is often modeled as a power law: 
\begin{equation}
%\begin{split}
f(M) = \frac{dN}{dM} \propto M^{\gamma}, 
%\end{split}
\end{equation}
in a differential form, where $M$ is the molecular gas mass and $N$ is the number of molecular clouds.

In the case of GMCs in the inner disk of the Milky Way, $\gamma$ is know to be in the range from $-1.6$ (\cite{Williams1997}) to $-1.9$ (\cite{Heyer2001}). However, recent statistical studies of GMCs in the outer disk of the Milky Way and local galaxies such as LMC, M33, M51\textcolor{black}{, and NGC 4526} suggest that the shape of the GMC mass function can vary depending on the environment, i.e., the regions where GMCs reside (e.g., \cite{Rosolowsky2005b}; \cite{Wong2011}; \cite{Gratier2012}; \cite{Colombo2014} \textcolor{black}{; \cite{Utomo2015}}), and theoretical studies of such environmental dependence will bring insights into the formation processes of molecular clouds (e.g., \cite{Inutsuka2015}). Because all of these studies are based on $^{12}$CO measurements in quiescent and relatively moderate star-forming galaxies, it is of interest to test whether the GMC mass function derived from $^{13}$CO in more actively star-forming regions gives a similar slope or not.

Figure \ref{fig:8} shows the derived LTE mass function in an integral form for the identified 187 GMCs in the central $\sim$4 kpc region of NGC 1068.
Note that the mass threshold of our GMC sample, $\sim1.7 \times 10^4$ $M_\odot$, corresponds to $>$8 $\sigma$ in the $^{13}$CO data cube. Therefore we will be comfortably safe to discuss the slope of the mass function above $\sim 10^4$ $M_\odot$ here. 

We find that the overall shape of the $^{13}$CO-based GMC mass function in NGC 1068 is similar to those in the Milky Way and local galaxies based on $^{12}$CO; i.e., it can be expressed as a power law with a cut-off at the high mass end ($M > 10^{7} M_\odot$). Here we adopt a truncated power law, 
\begin{equation}
N(M'>M) = N_0 \left[\left(\frac{M}{M_0}\right)^{\gamma+1}-1\right], 
\end{equation}
where $N_0$ is the number of clouds more massive than $2^{1/(\gamma+1)}M_0$, which is the mass where the distribution deviates from a single power law (e.g., \cite{Mckee1997}; \cite{Wong2011}; \cite{Colombo2014}). 
The fit to the data (for $M>10^5 M_\odot$) gives the slope $\gamma=-1.25\pm0.07$, the maximum mass $M_0 = (5.92 \pm 0.63) \times 10^7$ $M_\odot$, and 
%%the number of GMCs at the maximum mass 
$N_0$ = $54.4 \pm 28.2$. 

The derived slope of the mass function for GMC ($M>10^5 M_\odot$) in the disk of NGC 1068, $\gamma = -1.25\pm0.07$, is significantly shallower compared with those of Galactic GMCs ($\gamma = -1.6$: \cite{Williams1997}, $\gamma = -1.9$: \cite{Heyer2001}) and the disk GMCs in M51 ($\gamma = -1.8$ to $-2.5$, \cite{Colombo2014}) and NGC 300 ($\gamma = -1.8$, \cite{Faesi2016}), 
indicating that the more massive clouds dominate the mass budget in NGC 1068 than these disk GMCs. 
Furthermore, although we observed a cut-off of the mass function at the high mass end, 
as in the case of GMCs in the Milky Way and other local galaxies such as M51, the measured maximum mass, $M_0 \sim 6 \times 10^7 M_\odot$, is one order of magnitude larger than that in the nuclear bar region of M51 (a shallow slope, $\gamma \sim -1.3$, similar to NGC 1068, and a sharp high-mass cut-off above $M_0 \sim 5.5 \times 10^6$ $M_\odot$ are reported
%%in the nuclear bar region of M51, 
\cite{Colombo2014}). 
Presumably the presence of such more massive GMCs in NGC 1068 than \textcolor{black}{the Milky Way} and M51 could be a driver of intense starburst there; 
the elevated star formation activities in the starburst ring of NGC 1068 ($\Sigma_{\rm SFR} \sim 1 - 2$ $M_\odot$ yr$^{-1}$ kpc$^{-2}$, \cite{Tsai2012}), which is significantly larger than those in the spiral arms of M51 (mostly $\Sigma_{\rm SFR}$ $\sim$ a few $\times 10^{-1}$ $M_\odot$ yr$^{-1}$ kpc$^{-2}$ or less, e.g., \cite{Watanabe2016}), could be caused by such \textcolor{black}{a} difference in the GMC mass function.

\subsection{Spatial distribution of virial parameters}

Figure \ref{fig:VirialParameter-on-13COmap} displays the positions of the GMCs along with their virial parameters. 
We find that the supercritical GMCs,  i.e., $M_{\rm ^{13}CO}/M_{\rm vir} > 1$, are preferentially found on the starburst ring, 
while the subcritical GMCs are mainly located both inside and outside of the ring.
These are consistent with the trend of GMAs in M83 based on $^{12}$CO($J$=3--2) which trace dense gas (\cite{Muraoka2009}).
It is remarkable that these trends have also been found by observations of $^{13}$CO($J$=1--0), which is an optically thin
%%, ergo a more faithful 
molecular mass tracer, at GMC scales.
Such environmental difference \textcolor{black}{depending on the location of GMCs, i.e., whether {\it on-arm} or {\it inter-arm},} has been reported by \textcolor{black}{$^{12}$CO-based} statistical studies of GMCs in IC 342 (\cite{Hirota2011}), M51 (\cite{Colombo2014}) and \textcolor{black}{local spiral galaxies such as NGC 6946 and M101 (e.g., \cite{Rebolledo2015})}. 
%%This
\textcolor{black}{Our result} suggests that GMCs outside the starburst ring \textcolor{black}{in NGC 1068} are not gravitationally bound, giving an insight into formation mechanism of GMCs especially in the inter-arm regions.

%%%%%%%%%%%%%%%%%%%
%%% section 4.4 %%%
%%%%%%%%%%%%%%%%%%%

\subsection{GMC-scale variation of molecular line properties and their link to the larger scale environments in NGC 1068}
\label{sec:GMC_properties}

\begin{figure*}[H]
  \begin{center}
   \FigureFile(80mm,100mm){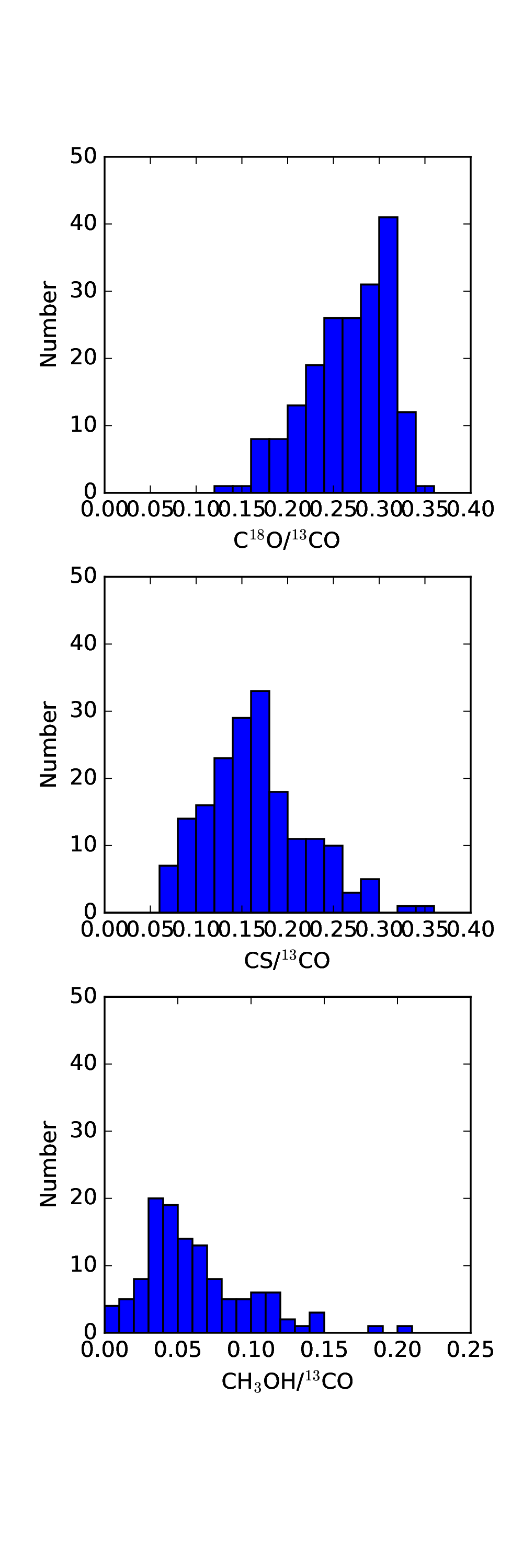}
    %%% \FigureFile(width,height){filename}
  \end{center}
  \caption{Histograms of the measured {integrated intensity} ratios of the identified GMCs.}\label{fig:5}
\end{figure*}

\begin{figure*}[H]
  \begin{center}
   \FigureFile(120mm,120mm){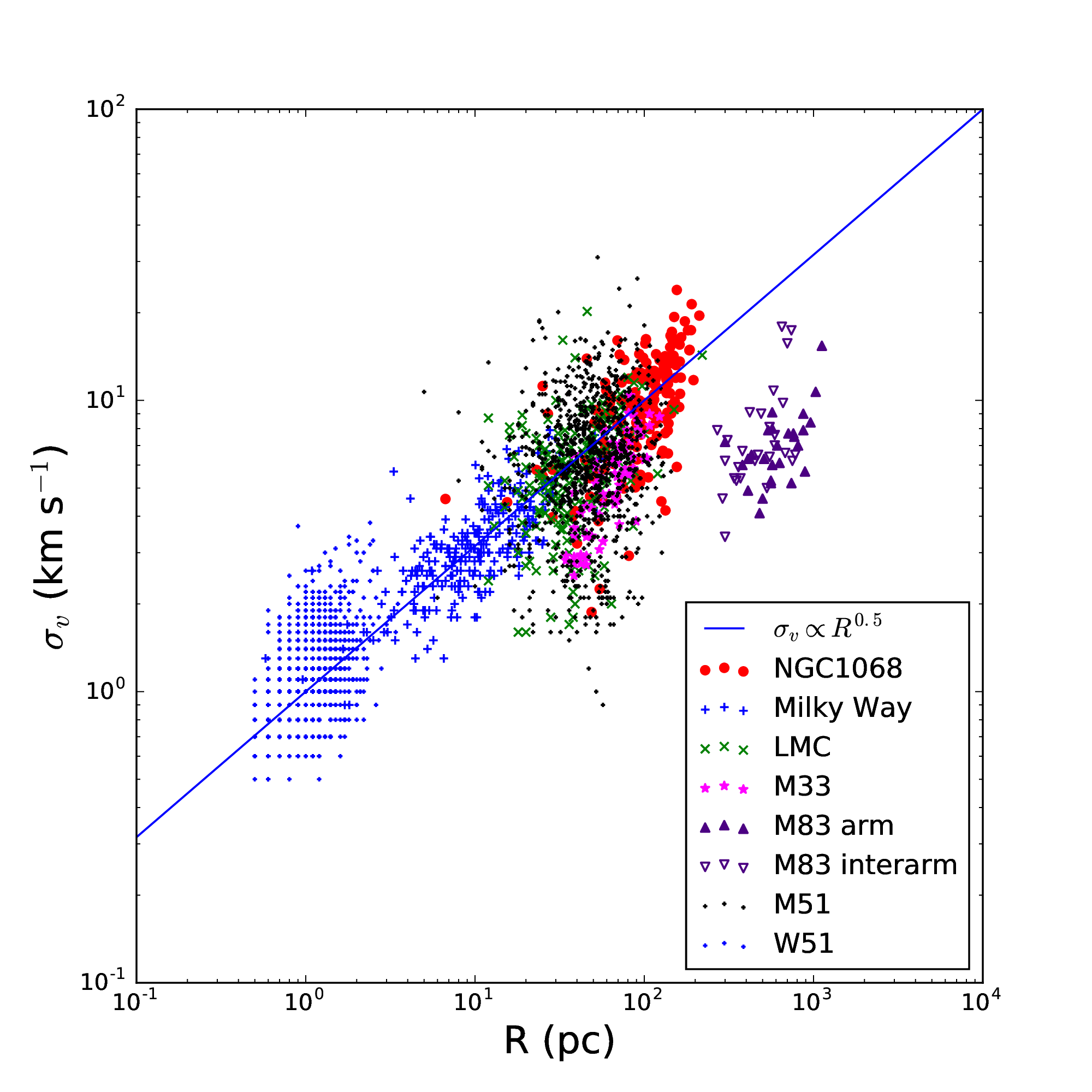}
    %%% \FigureFile(width,height){filename}
  \end{center}
  \caption{The correlation between size and velocity width of the identified GMCs based on the $^{13}$CO($J$=1--0) cube in NGC 1068, together with molecular clouds in the Milky Way (\cite{Sanders1985}, based-on $^{12}$CO($J$=1--0)), cores/clumps in the Galactic star-forming region W51 (\cite{Parsons2012}, $^{13}$CO($J$=3--2); note that the data points are quantized due to the original data), and GMCs/GMAs in local galaxies including LMC (\cite{Fukui2008}, $^{12}$CO($J$=1--0)), M33 (\cite{Onodera2012}, $^{12}$CO($J$=1--0)), M51 (\cite{Colombo2014}, $^{12}$CO($J$=1--0)) and M83 (\cite{Muraoka2009}, $^{12}$CO($J$=3--2)). The diagonal line indicates the Larson's raw, $\sigma_v \propto R^{0.5}$ (\cite{Larson1981}). }\label{fig:6}
\end{figure*}

\begin{figure*}[H]
  \begin{center}
   \FigureFile(150mm,150mm){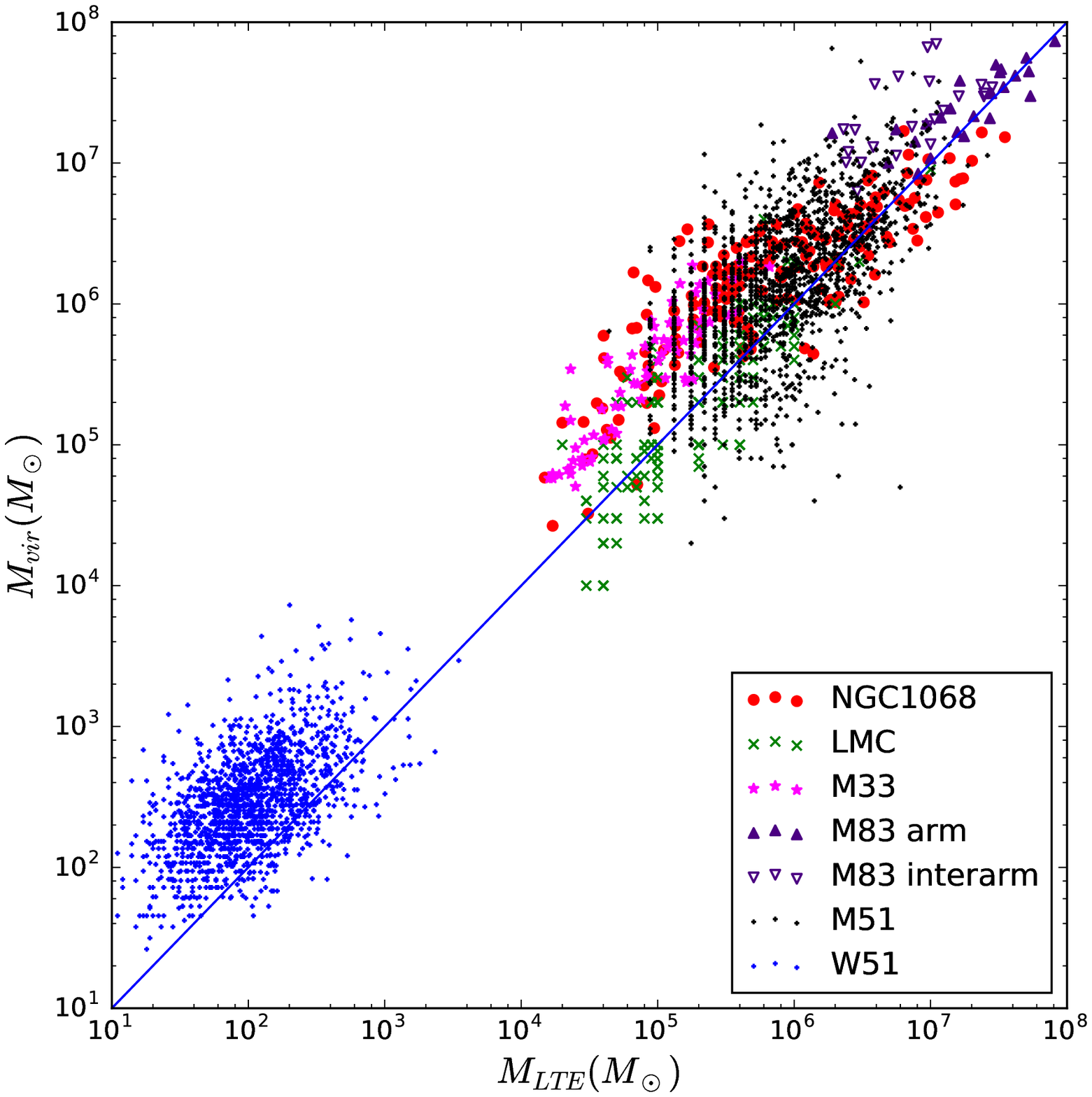}
    %%% \FigureFile(width,height){filename}
  \end{center}
  \caption{The correlation between molecular gas mass (LTE mass) and virial mass of the identified GMCs based on the $^{13}$CO($J$=1--0) in NGC 1068, together with molecular cores/clumps in W51 (\cite{Parsons2012}, $^{13}$CO($J$=3--2); note that the data points are quantized due to the original data), and GMCs/GMAs in local galaxies including LMC (\cite{Fukui2008}, $^{12}$CO($J$=1--0)), M33 (\cite{Onodera2012}, $^{12}$CO($J$=1--0)), M51 (\cite{Colombo2014}, $^{12}$CO($J$=1--0)) and M83 (\cite{Muraoka2009}, $^{12}$CO($J$=3--2)). The diagonal line shows $M_{\rm LTE} = M_{\rm vir}$. }\label{fig:7}
\end{figure*}

\begin{figure*}[H]
  \begin{center}
   \FigureFile(150mm,150mm){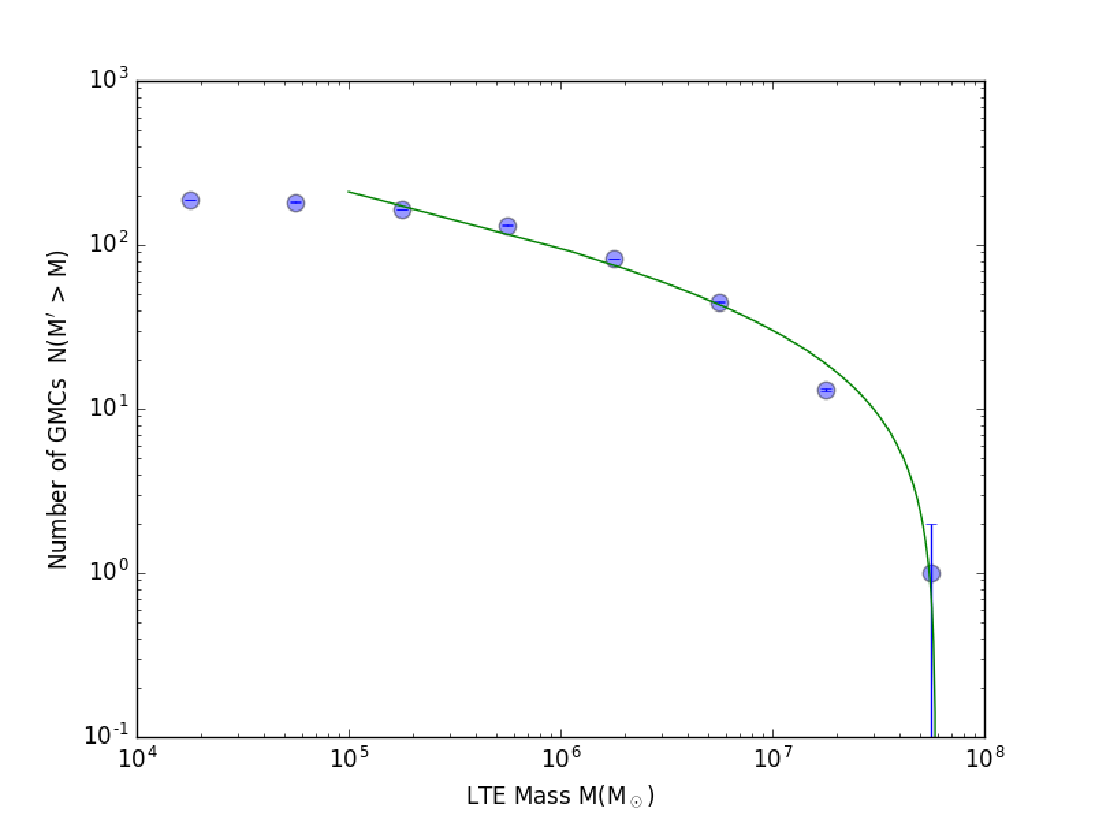}
  \end{center}
 \caption{Mass function of identified GMCs. The solid line is the fit to the data (for $M>10^5 M_\odot$) with the truncated power law function, $N(M'>M) = N_0 \left[\left(\frac{M}{M_0}\right)^{\gamma+1}-1\right]$, where the slope $\gamma=-1.25 \pm 0.07$, the maximum mass $M_0$ = $(5.92 \pm 0.63) \times 10^7$ $M_\odot$, and the number of GMCs at the maximum mass $N_0$ = $54.4 \pm 28.2$. }\label{fig:8}
\end{figure*}

\begin{figure*}[H]
  \begin{center}
   \FigureFile(150mm,150mm){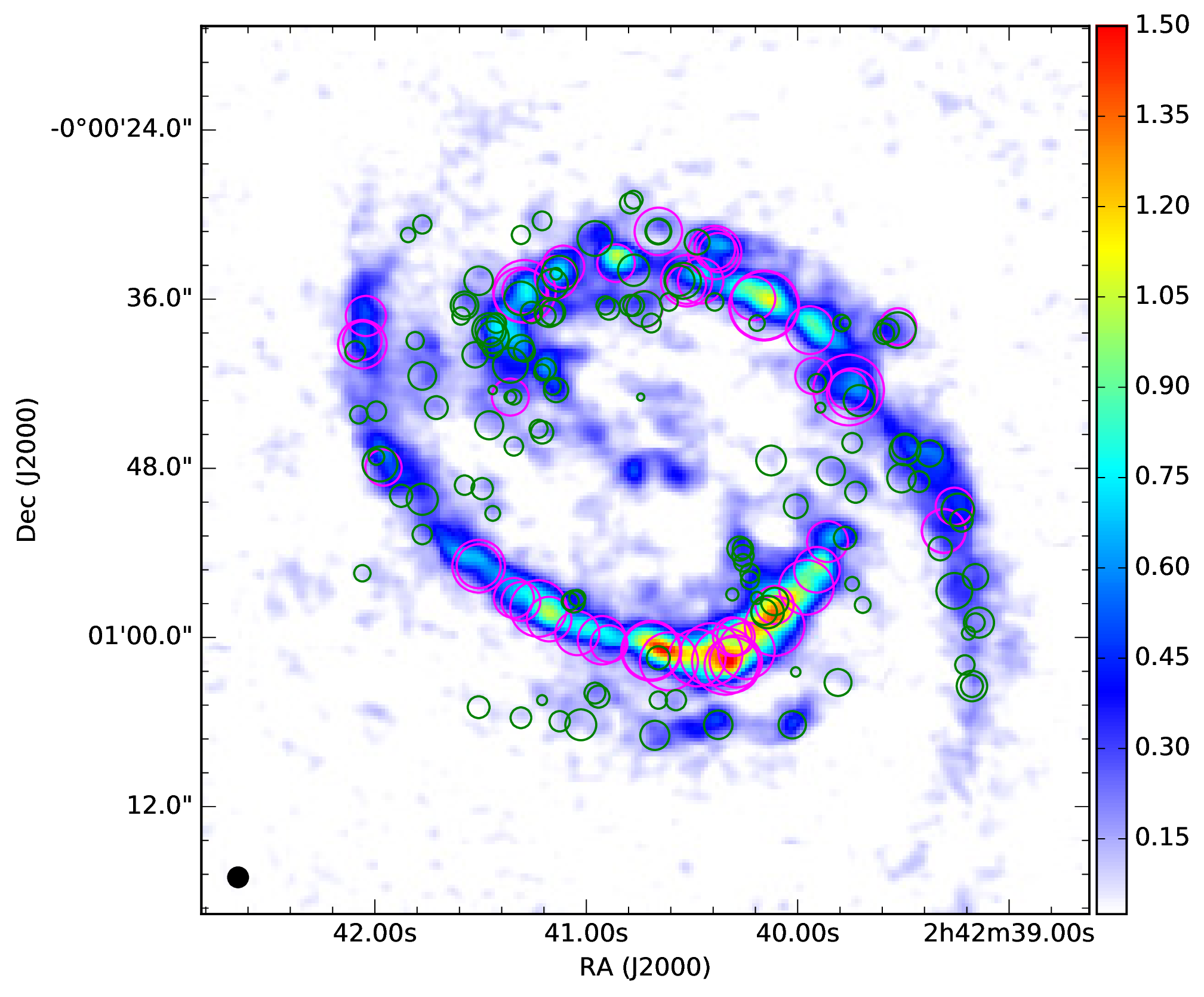}
    %%% \FigureFile(width,height){filename}
  \end{center}
  \caption{The position of the identified clouds and their virial parameter, $M_{\rm 13CO}/M_{\rm vir}$, superposed on the pseudo-color $^{13}$CO($J$=1--0) integrated intensity image. The size of circles is proportional to virial parameter of each cloud (from 0.04 to 3.9), and the color of circles indicate their boundness; magenta and green circles are GMCs with virial parameter larger than \textcolor{black}{1} (i.e., supercritical) and smaller than \textcolor{black}{1} (subcritical), respectively. 
}\label{fig:VirialParameter-on-13COmap}
\end{figure*}

\begin{figure*}[H]
  \begin{center}
   \FigureFile(180mm,180mm){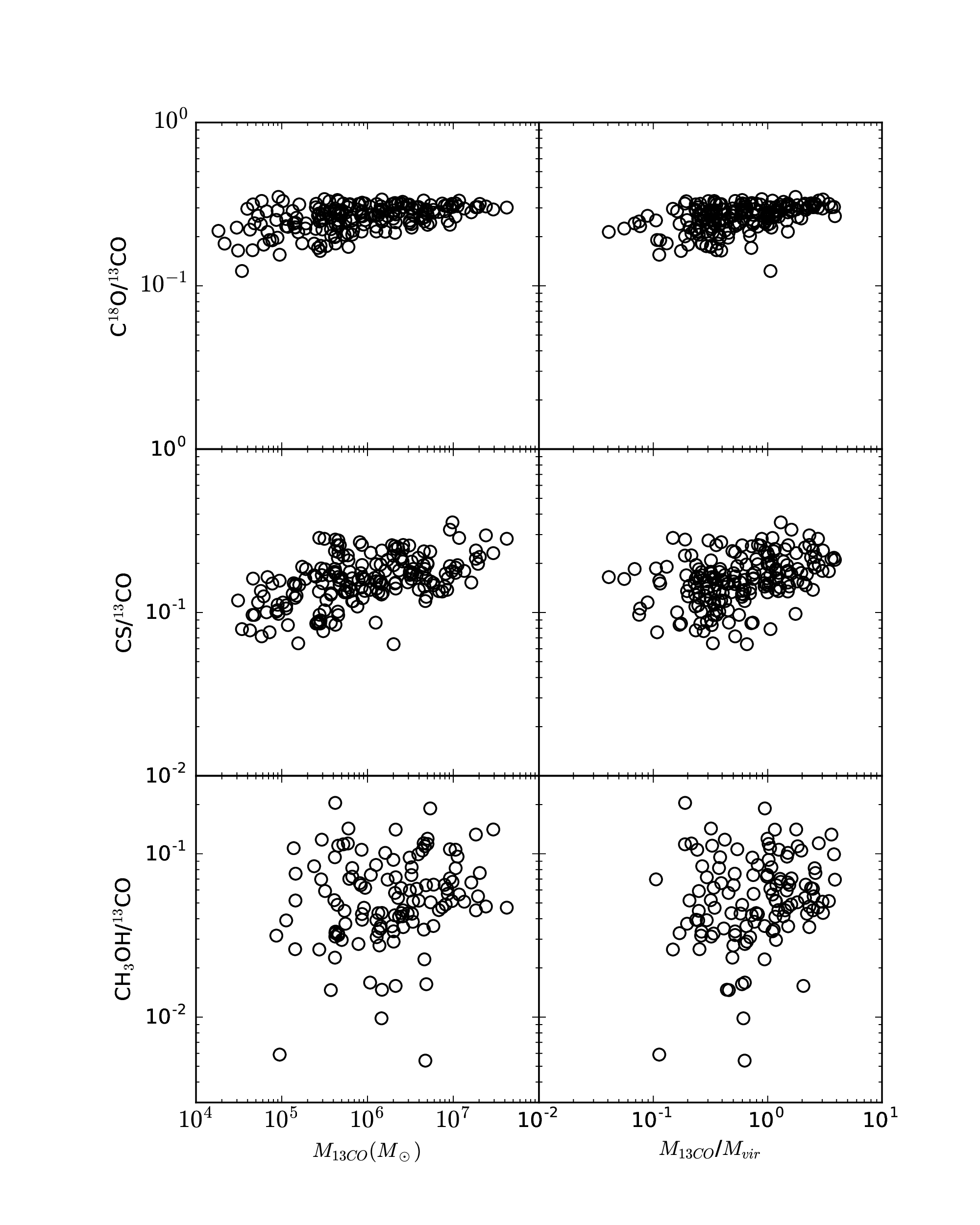}
    %%% \FigureFile(width,height){filename}
  \end{center}
  \caption{The correlation between molecular gas mass (LTE mass) from $^{13}$CO and {integrated intensity} ratio (left 3 panels)  and that between virial parameter and line ratio (right 3 panels).}\label{fig:9}
\end{figure*}

\begin{figure*}[H]
  \begin{center}
   \FigureFile(90mm,180mm){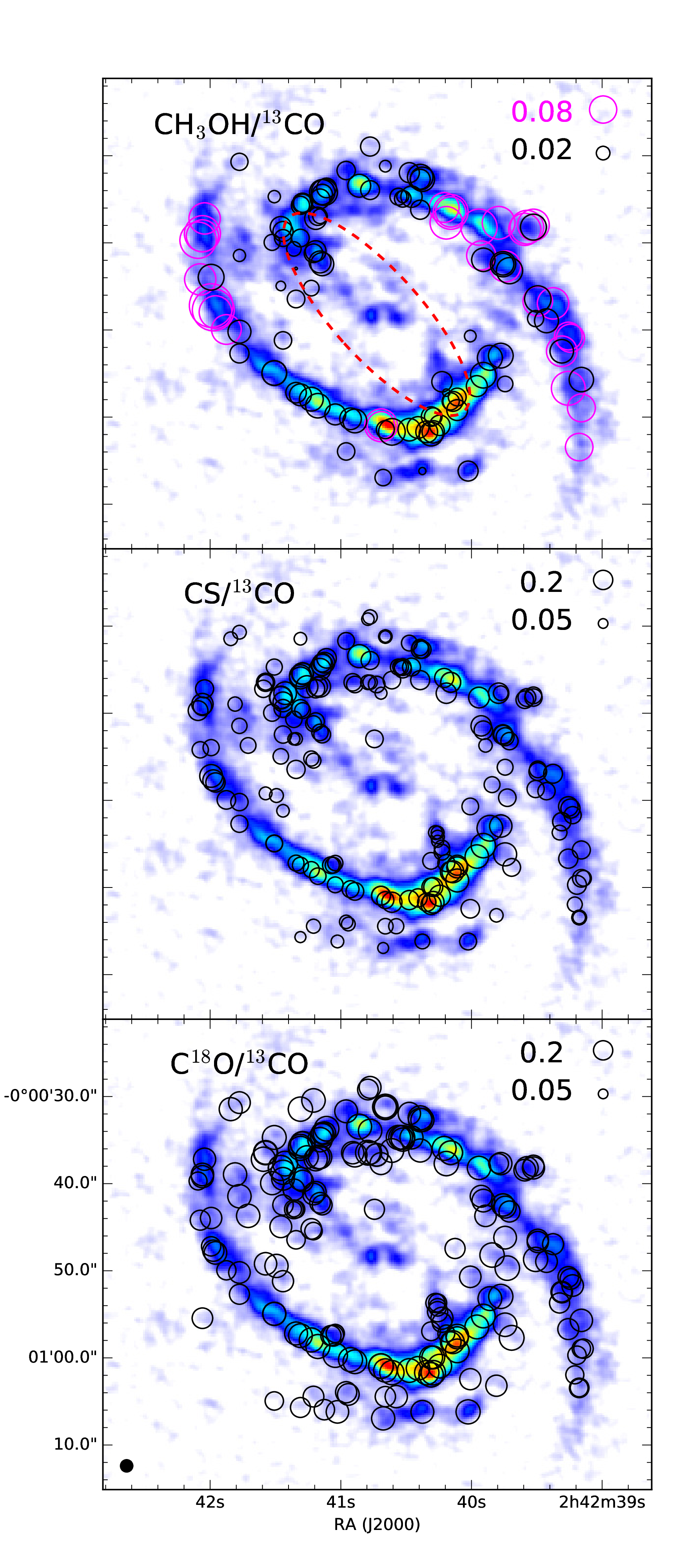}
    %%% \FigureFile(width,height){filename}
  \end{center}
  \caption{Spatial distribution of CH$_3$OH/$^{13}$CO, CS/$^{13}$CO, and C$^{18}$O/$^{13}$CO integrated intensity ratios of individual clouds. The size of each circle is proportional to the line ratios of C$^{18}$O, CS, CH$_3$OH with respect to $^{13}$CO. In the top panel, GMCs with the enhanced methanol emission (CH$_3$OH/$^{13}$CO ratios larger than 0.08) are shown in magenta circles. The red dashed ellipse shows the nuclear stellar bar seen in the near-infrared (P.A.= 48$^\circ$ and the semi-major axis = $16''$, \cite{Scoville1988}). The CH$_3$OH/$^{13}$CO ratios are smallest around the bar-end, whereas the GMCs with elevated CH$_3$OH/$^{13}$CO ratios preferentially reside in the upper stream side of the spiral arms. Note that gas particles go around counter-clockwise.}\label{fig:10}
\end{figure*}

In section \ref{sec:M_13CO-Mvir}, we have shown that the histograms of C$^{18}$O/$^{13}$CO and CS/$^{13}$CO \textcolor{black}{integrated intensity ratios (hereafter we refer to as ratios)} have single peaked distribution, while the histogram of CH$_3$OH/$^{13}$CO ratios exhibits widely scattered distribution (\textcolor{black}{figure} \ref{fig:5}). In order to test if these molecular line ratios follow any trend with other physical quantities, these line ratios are plotted against the molecular gas masses and virial parameters of each GMC (figure \ref{fig:9}). We find C$^{18}$O/$^{13}$CO ratios are mostly uniform regardless of the molecular gas masses and virial parameters (the \textcolor{black}{top panels} of figure \ref{fig:9}), whereas CS/$^{13}$CO ratios show a positive trend with the gas mass and virial parameters as shown in the \textcolor{black}{middle panels} of figure \ref{fig:9}. On the other hand, CH$_3$OH/$^{13}$CO ratios exhibit different behavior from C$^{18}$O/$^{13}$CO and CS/$^{13}$CO ratios; they show significantly larger scatters and no clear systematic tendency was found. 
We then display the spatial distributions of these molecular line ratios in figure \ref{fig:10}, in order to see any connection exists between the GMC-scale molecular line ratios and larger scale environments such as spiral arms and circumnuclear starburst regions. In this figure, the size of each circle is proportional to the ratios of each molecule with respect to $^{13}$CO. It is very clearly displayed that the diameters of circles representing the C$^{18}$O/$^{13}$CO and CS/$^{13}$CO ratios are almost uniform across the starburst ring, as expected from the narrow variation of the ratio (figure \ref{fig:5}). 
On the other hand, in the top panel of figure \ref{fig:10}, the circle diameters vary significantly from cloud to cloud in the  CH$_3$OH/$^{13}$CO map. The CH$_3$OH/$^{13}$CO ratios are smallest around the bar-end, and become larger along the spiral arms.
That is, importantly, we find systematic spatial variations of the CH$_3$OH/$^{13}$CO ratios across the starburst ring
of NGC 1068 at a GMC scale. 

Here we discuss the possible physical and chemical processes which govern the observed behaviors of these molecular line ratios below.

%% section 4.4.1
\subsubsection{C$^{18}$O/$^{13}$CO ratios: uniformity among GMCs}
\label{sec:line_ratios_vs_gasmass}

The C$^{18}$O/$^{13}$CO ratios (the \textcolor{black}{top panels} of figure \ref{fig:9}) are found to be mostly uniform regardless of the molecular gas mass and virial parameter, as expected from figure 5, demonstrating that the C$^{18}$O/$^{13}$CO ratios are almost within a factor of 2. The average value of the C$^{18}$O/$^{13}$CO ratios is 
%$0.27 \pm 0.046$ 
$0.27 \pm 0.05$ 
as summarized in table 4. This value is similar to other high-resolution (GMC or GMA-scale) extragalactic C$^{18}$O/$^{13}$CO ratios such as in NGC 6946 (\cite{Meier2004}), IC 342 (\cite{Meier2005}) and M 51 (\cite{Watanabe2016}); see also the top panel of figure 12, summarizing these extragalactic, GMC/GMA-scale C$^{18}$O/$^{13}$CO ratio measurements. Note that the 10-pc scale measurements of C$^{18}$O/$^{13}$CO ratios of GMCs in LMC reveal significantly lower values than these galaxies ($<0.1$; \cite{Nishimura2016}). This could presumably be due to the selective far-UV dissociation of C$^{18}$O (\cite{vanDishoeck1988}), which is indeed observed in the Orion-A giant molecular cloud (\cite{Shimajiri2014}). However, the observed spatial uniformity of the C$^{18}$O/$^{13}$CO ratios indicate that such isotope-selective photo-dissociation, which is expected to be significant in GMCs on spiral arms (where intense starbursts occur) and less significant in inter-arm GMCs, plays only a negligible role in these measured GMC-scale C$^{18}$O/$^{13}$CO ratios. 

Because both the C$^{18}$O and $^{13}$CO $J$=1--0 lines are considered to be optically thin, the ratio can be regarded as the [C$^{18}$O]/[$^{13}$CO] fractional abundance ratio in the 0-th approximation. In \textcolor{black}{the Milky Way}, the [C$^{18}$O]/[$^{13}$CO] abundance ratio is known to be $\sim$1/7 -- 1/8 (\cite{Henkel1993}). Owing to the different production processes of C$^{18}$O (comes from short-lived massive stars during an early phase of a starburst event,  \cite{Prantzos1996}) and $^{13}$CO (produced later on in intermediate-mass stars, \cite{Wilson1992}; \cite{Meier2004} and references therein), it is suggested that GMCs preferentially forming massive stars can display elevated C$^{18}$O abundance with respect to $^{13}$CO (\cite{Henkel1993}; \cite{Meier2004}; \cite{Konig2016}). The measured C$^{18}$O/$^{13}$CO ratio or [C$^{18}$O]/[$^{13}$CO] abundance ratio of 0.27 or $\sim$1/4, is larger than that in the Galactic GMCs by a factor of 2, and may be a reflection of the on-going dusty starburst activities in the observed spiral arms of NGC 1068. Surprisingly, an extremely elevated [C$^{18}$O]/[$^{13}$CO] abundance ratio of close to unity, has been reported in the gravitationally-magnified submillimeter galaxy (i.e., a dusty extreme starburst galaxy) SMM J2135-0102 at $z=2.325$, indicating the presence of intense star-formation preferentially biased to high-mass stars (\cite{Danielson2013}). Increasing the measurements of C$^{18}$O and $^{13}$CO lines in various environments will be crucial to understand and establish the [C$^{18}$O]/[$^{13}$CO] abundance ratio as a tool for studying the evolution of galaxies near and far in the ALMA era.

%% section 4.4.2
\subsubsection{CS/$^{13}$CO ratios: tracing a variation of dense gas fraction among GMCs?}

In the \textcolor{black}{middle panels} of figure \ref{fig:9}, we find a trend of CS/$^{13}$CO ratios with respect to the molecular gas mass and virial ratios; massive GMCs and supercritical GMCs tend to show higher CS/$^{13}$CO ratios, 
whereas smaller CS/$^{13}$CO ratios are found among less massive or subcritical GMCs. 

CS/$^{13}$CO ratios can be regarded as a proxy of dense gas fraction, because CS traces significantly denser molecular gas than $^{13}$CO owing to its higher critical density ($n_{\rm crit} \sim10^5$ cm$^{-3}$ for CS($J$=2--1)). 
Therefore, the relationship between these line ratios and GMC masses suggest that more massive GMCs form dense molecular gas more efficiently than less massive ones. 
This is in fact exactly what we found in M33, i.e., more massive GMCs tend to show higher $^{12}$CO($J$=3--2)/$^{12}$CO($J$=1--0) line ratios, 
which can be interpreted as another indicator of dense gas fraction for relatively quiescent disk regions (\cite{Onodera2012}).

%% section 4.4.3 
\subsubsection{Spatial variation of CH$_3$OH$/^{13}$CO ratios: caused by mild shocks along spiral arms?}

In figure \ref{fig:10}, we find that the CH$_3$OH/$^{13}$CO ratios are smallest around the bar-end, and become larger along the spiral arms, i.e., we find systematic spatial variations of the CH$_3$OH/$^{13}$CO ratios across the starburst ring of NGC 1068 at a GMC scale.
In fact, similar tendencies have been reported in the central regions of IC 342 and Maffei2; the distribution
of CH$_3$OH is spatially anti-correlated with those of $^{13}$CO and C$^{18}$O (\cite{Meier2005}; \cite{Meier2012}), 
giving a coherent view with our findings.

Here, we discuss the origin of the observed systematic spatial variation of CH$_3$OH/$^{13}$CO ratios 
across the starburst ring in NGC 1068. If we consider the direction of the galactic rotation 
in the disk of NGC 1068 (gas particles go around counter-clockwise), we can say that 
the CH$_3$OH/$^{13}$CO ratios become larger in the up-stream side, and then are decreasing in the down-stream direction. 
Two possibilities are: (1) GMCs with different line ratios are produced locally, and 
(2) GMCs with high line ratios in the up-stream side of spiral arms travel along the arms, 
and then the line ratios are gradually decreasing. 

The chemical timescale of CH$_3$OH formation and destruction is known to be $10^5$ years (\cite{Nomura2004}). 
The distance of the starburst ring/spiral arms from the nucleus is approximately 20$^{\prime\prime}$ - 30$^{\prime\prime}$ (1.6 - 2.4 kpc), 
and therefore the dynamical timescale (i.e., the time necessary for one rotation) is 5 - 7 $\times 10^7$ years
if we adopt a circular rotation velocity of 200 km s$^{-1}$ at this radius (\cite{Schinnerer2000}). 
Two distinct GMCs with high and low CH$_3$OH/$^{13}$CO line ratios are separated by $\sim$0.5 rotation, 
and it takes approximately $\sim 10^7$ years to travel between these two positions. This time scale is significantly 
longer than the chemical formation/destruction time scale of $10^5$ years. We therefore conclude that 
scenario (1) is more likely than (2), i.e., GMCs with different chemical properties are 
locally formed in different places. 

What is the cause of such local GMC variation in methanol intensity with respect to $^{13}$CO? 
One possible cause is the local difference of either shock strength or dust temperature. 
CH$_3$OH is suggested to be a tracer of mild shock, because its gas-phase abundance is enhanced in a moderately 
intense shock whereas it is destroyed under the strong shock environments (\cite{Viti2011}). 
Shock strengths should be stronger in the bar-end than spiral arms because  the bar-end is the orbit-crowding 
region of two types of gas motion, $x_1$ and $x_2$ orbits (e.g. \cite{Athanassoula1999}). 
Although shock is expected 
along the spiral arms (spiral shock waves, \citet{Fujimoto1968}, \citet{Roberts1969}, see also \citet{Baba2013} for a recent theoretical view on such spiral shocks), 
it can be rather milder than that in the bar-end. 
It is expected that the shock in the bar-end 
is so strong that methanol molecules are destroyed, 
although no significant SiO emission is reported around the bar-end of NGC 1068
(\cite{GarciaBurillo2010}), in spite of the fact that SiO is a tracer of strong shock.

\begin{figure*}[H] 
  \begin{center}
   \FigureFile(180mm,180mm){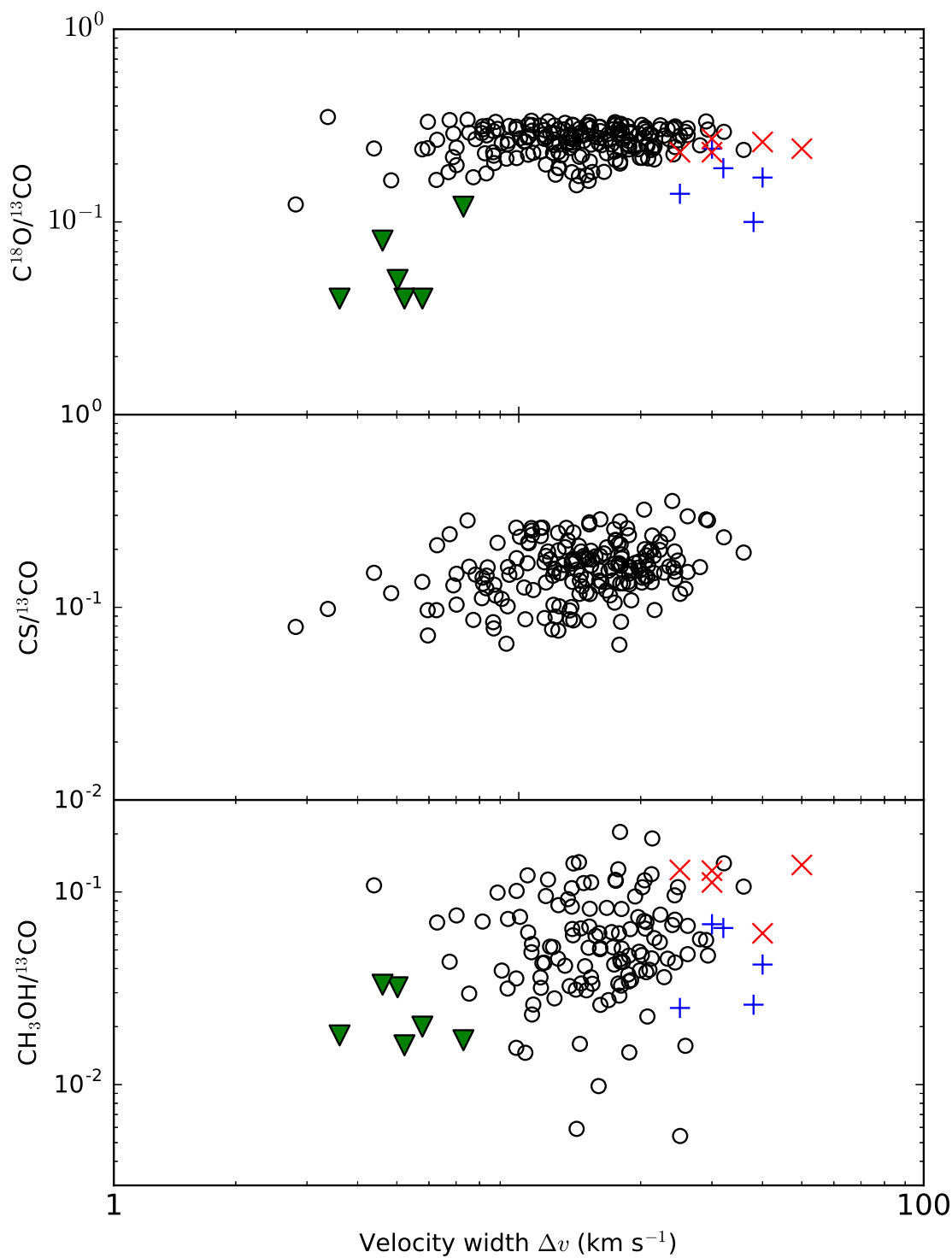}
      %%% \FigureFile(width,height){filename}
  \end{center}
  \caption{The {integrated intensity} ratios of C$^{18}$O, CS, CH$_3$OH to $^{13}$CO, plotted against velocity widths of GMCs in NGC 1068 (open circles). Blue pluses, red crosses, and green triangles are those in IC 342 (\cite{Meier2005,Meier2012}, $\sim 100$ pc), M 51(\cite{Watanabe2016}, $\sim 200$ pc), and LMC (\cite{Nishimura2016}, $\sim 10$ pc, upper limit), respectively. }\label{fig:12}
\end{figure*}

Can we see a possible link between the shock strengths and methanol abundance enhancement? One of the hint can be drawn from the comparison of the CH$_3$OH/$^{13}$CO ratios and velocity widths of GMCs, because the velocity widths are dominated by internal turbulent velocities (although systematic velocity gradient such as disk rotation should be appropriately modeled for accurate analysis). 
In figure \ref{fig:12}, the CH$_3$OH/$^{13}$CO integrated intensity ratios of GMCs are plotted against the velocity widths of GMCs in the central $\sim$4 kpc region of NGC 1068, along with the C$^{18}$O/$^{13}$CO and CS/$^{13}$CO ratios. The CH$_3$OH/$^{13}$CO ratios in local galaxies including IC 342 (\cite{Meier2005}; \cite{Meier2012}, $\sim$100 pc resolution), M 51 (\cite{Watanabe2016}, $\sim$200 pc), and LMC (\cite{Nishimura2016}, $\sim$10 pc, upper limits only) are also shown for comparison. We find no clear trend between CH$_3$OH/$^{13}$CO ratios and velocity widths of clouds, implying that the correlation between shock and methanol abundance is not very obvious; presumably, a putative correlation may be weakened due to the fact that strong shock is not preferable for methanol. In fact, no clear correlation between star formation activities and enhancement of methanol was found in our previous ALMA study of NGC 1068 (\cite{Takano2014}), implying that strong shocks caused by successive supernovae in active starburst regions are not responsible for the enhanced methanol emission. 

\begin{figure*}[H]
  \begin{center}
   \FigureFile(180mm,180mm){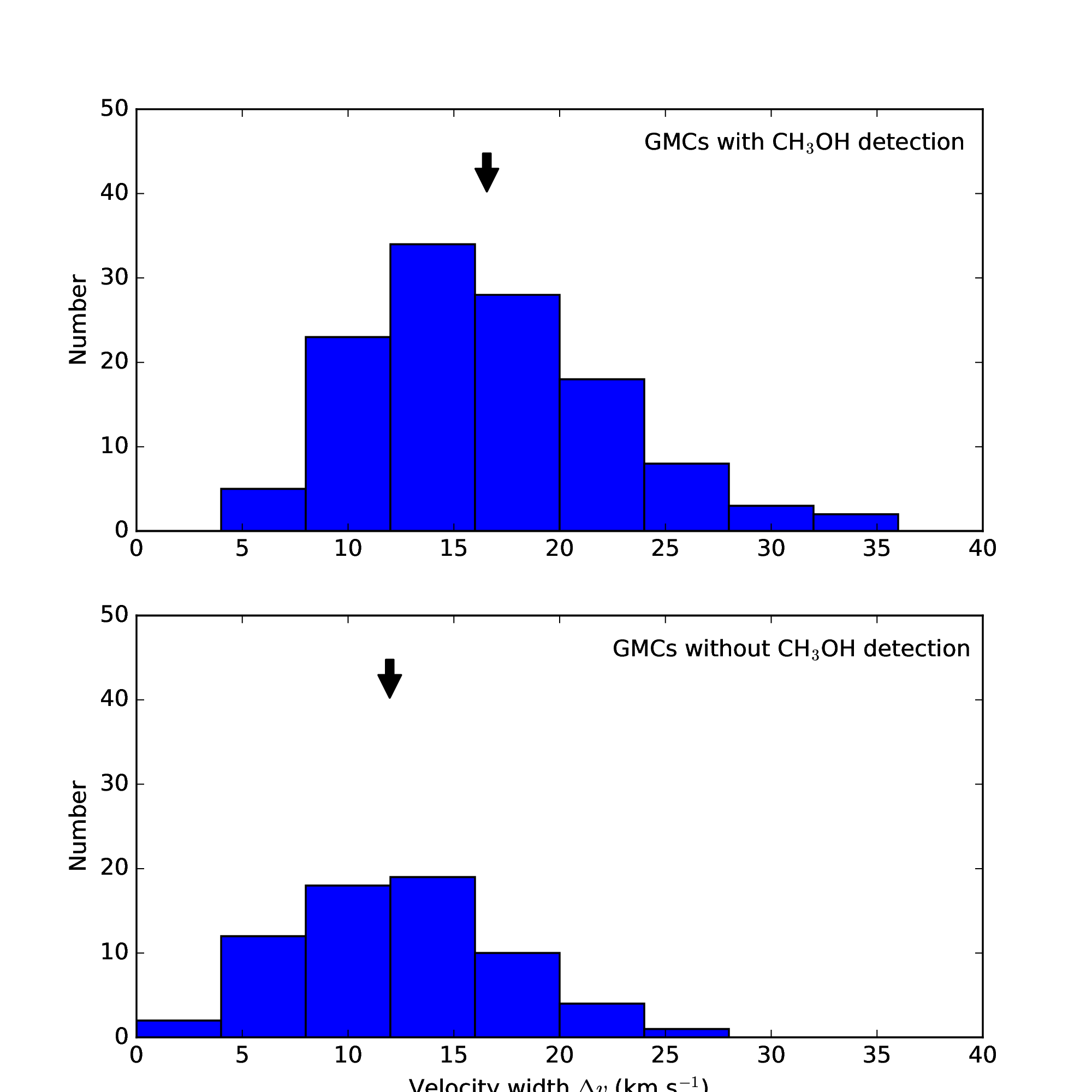}
      %%% \FigureFile(width,height){filename}
  \end{center}
  \caption{(top) Histogram of line widths of GMCs with significant detection of methanol. The average value is indicated by an arrow. (bottom) The same as the top panel but for GMCs with no methanol emission. The comparison of these two histograms suggests that the GMCs with methanol emission tend to have larger velocity line widths than those without methanol emission. A two-sample Kolmogorov Smirnov test gives $p$ value of less than $10^{-5}$, i.e., a hypothesis that these two samples are originated from the same distribution is rejected with a $>99.9$\% significance.  }\label{fig:13}
\end{figure*}

Another test to address the link between methanol abundance and shock is to compare the velocity widths of GMCs with and without methanol detection; figure \ref{fig:13} shows histograms of line widths of GMCs with and without detection of methanol. The mean and standard deviation are 16.56 km s$^{-1}$ and 5.61 km s$^{-1}$ for methanol-detected GMCs, and 11.97 km s$^{-1}$ and 4.87 km s$^{-1}$ for methanol-lacking GMCs, respectively, suggesting that methanol-detected GMCs seem to have larger velocity widths. In fact, 
a two-sample Kolmogorov-Smirnov test gives a $p$ value of less than $10^{-5}$, 
 i.e., a hypothesis that these two samples are originated from the same distribution is rejected with a $>99.9$\% significance.
The comparison of these two histograms suggests that the GMCs with methanol emission tend to have larger velocity line widths 
 than those without methanol emission. 
This implies that methanol emission originates from shocks which are ubiquitous in spiral arms, regardless of the intensity of star formation activities there. We therefore suggest that spiral shocks associated with spiral arms are responsible for the presence of methanol emission in the spiral arms in NGC 1068 and other galaxies such as M 51. 

Next, we consider the role of different temperatures as a driver of methanol abundance variation 
among the GMCs. It is know\textcolor{black}{n} that the yield of methanol molecule becomes maximum 
around the dust temperature of 15 K, and it decreases above the higher dust temperature ($\geq$ 20K; \cite{Watanabe2003}). 
This means that only the molecular (and dust) clouds staying at a low temperature 
($\leq$ 20-30K) for a chemical time scale can enhance the abundance of CH$_3$OH. 
Therefore, we can hypothesize that the temperature of gas and dust around the bar-end is significantly high 
($\geq$ 30K) and that is the cause of low CH$_3$OH/$^{13}$CO ratio. It is natural to expect that 
massive star-formation heats the ISM; in fact, the measured star-formation rates 
in the bar-end region is $\sim$2 times higher than those in the spiral arms (\cite{Tsai2012}).

With these pieces of evidence and implications, we can draw a scenario as follows. 
Methanol comes into gas phase by mild shock when molecular gas is accumulated and compressed along the spiral arms and bar-end. 
In the bar-end, however, more active massive 
star-formation occurs and therefore gas/dust temperature becomes higher than in the spiral arms. 
The production of methanol is then suppressed in the bar-end. In this way, we observe 
low CH$_3$OH/$^{13}$CO ratio in the bar-end, whereas elevated CH$_3$OH/$^{13}$CO ratios were found along the spiral arms. 

In order to verify this scenario quantitatively, 
we need to determine the spatial distribution of gas and dust temperatures along the starbursting spiral arms, which is not yet available.
Future ALMA observations will play a role in conducting such studies.

\section{Summary}

We present a statistical study of physical and chemical properties of giant molecular clouds (GMCs) in the central $1'$ (4.2 kpc) region of NGC 1068, based on $1''.4$ (98 pc) resolution ALMA  observations of $^{13}$CO($J$=1-0), C$^{18}$O($J$=1-0), CS($J$=2-1), and CH$_3$OH($J_K$=$2_K$-$1_K$). The observed region contains both active galactic nuclei and extended active star-forming regions, which do not exist in Galactic disks; this allows us to address the question of whether these extreme activities will have significant impact on the physical and chemical properties of ISM at a GMC scale.  

High sensitivity ALMA observations allowed us to produce 3-dimensional data cubes of these molecular lines with a high spectral resolution (1.5 km s$^{-1}$), which is essential for a study of GMC-scale molecular gas in the disk regions of galaxies. 

The major outcomes of this study are summarized as follows.

\begin{enumerate}

\item A high dynamic range (72) $^{13}$CO(1-0) spectral data cube, produced with a wide range of baseline lengths (5.5 k$\lambda$ -- 290 k$\lambda$) and multi-scale CLEAN, reveals detailed sub-structures of spiral arms and faint clouds in the inter-arm regions. 

\item We have identified 187 high significance GMCs from the $^{13}$CO(1-0) cube by employing the \verb|CLUMPFIND| algorithm with high level threshold (8 $\sigma$) and increment (4 $\sigma$) in order to ensure reliable decomposition of the $^{13}$CO(1-0) emission into individual clouds. This is one of the largest samples of GMCs in galaxies constructed based on $^{13}$CO(1-0) emission. 

\item We find that the cataloged GMCs tend to follow the known size to line-width relation of the Galactic (i.e., non-starbursting) GMCs. 
The molecular gas masses of GMCs $M_{\rm 13CO}$, derived from the $^{13}$CO data, range from $1.8 \times 10^4 ~ M_\odot$ to $4.2 \times 10^7 ~M_\odot$, and their ratios to the virial masses $M_{\rm vir}$, i.e., virial parameters, $M_{\rm 13CO}/M_{\rm vir}$, show that the super-critical GMCs (i.e., $M_{\rm 13CO}/M_{\rm vir}>1.1$) are preferentially found on the starburst ring, while the sub-critical GMCs are mainly located both inside and outside of the ring. 

\item A mass function of GMCs, $N(M'>M) = N_0 \left[\left(\frac{M}{M_0}\right)^{\gamma+1}-1\right]$, in the central $\sim$4 kpc region of NGC 1068 has been obtained for the first time at $\sim$ 100 pc resolution. We find the slope of the mass function $\gamma = -1.25 \pm 0.07$ for a mass range of $M \geq 10^5M_\odot$. This is shallower than the Galactic GMCs ($\gamma = -1.6$ to $-1.9$) and GMCs in the disk regions of M51 ($\gamma = -1.8$ to $-2.5$) and NGC 300 ($\gamma = -1.8$), presumably caused by the presence of more massive clouds in the massive end of the mass function in NGC 1068. A large maximum mass, $M_0 = (5.92 \pm 0.63) \times 10^7$ $M_\odot$, also supports this view. 

\item The observed C$^{18}$O($J$=1--0)/$^{13}$CO($J$=1--0) intensity ratios are found to be fairly uniform \textcolor{black}{(0.27 $\pm$ 0.05)} among the identified GMCs. In contrast, the CH$_3$OH/$^{13}$CO ratios exhibit striking variation across the disk, with the smallest values around the bar-end ($<$ 0.03), and becoming larger along the spiral arms ($\sim$ 0.1--0.2). 

\item Conceivable causes which may govern the variation of CH$_3$OH/$^{13}$CO ratios across the disk of NGC 1068 have been investigated. We find that GMCs with methanol emission tend to have systematically larger velocity widths than those without methanol emission. We suggest that relatively weak shocks, which are ubiquitous in disk regions, i.e., spiral shocks, are responsible for the enhancement of the CH$_3$OH/$^{13}$CO ratios of GMCs in the disk of NGC 1068. 

\end{enumerate}

Finally, we stress that our study demonstrates the power of ALMA which relates the GMC-scale physical and chemical properties to the larger-scale environment of galaxies. Chemical properties revealed by ALMA  
will be an indispensable tool to study the physical processes of galaxy evolution near and far 
by its unprecedentedly high angular resolution and sensitivity. 

\bigskip

\begin{ack}
\textcolor{black}{We thank the anonymous referee for %%very kind and 
helpful comments for improving the paper.}
{Data analysis were in part carried out on common use data analysis computer system at the Astronomy Data Center, ADC, of the National Astronomical Observatory of Japan (NAOJ).}
This paper makes use of the
following ALMA data: ADS/JAO.ALMA\#2012.1.00060.S.
ALMA is a partnership of ESO (representing its member
states), NSF (USA), and NINS (Japan), together with NRC
(Canada), NSC and ASIAA (Taiwan), and KASI (Republic of Korea), in cooperation with
the Republic of Chile. The Joint ALMA Observatory is
operated by ESO, AUI/NRAO, and NAOJ. 
\textcolor{black}{T.T. was supported by
the ALMA Japan Research Grant of NAOJ Chile Observatory,
NAOJ-ALMA-0156. 
K.K. acknowledges support from JSPS KAKENHI grant number 25247019.}
\end{ack}

\begin{longtable}{ccccccc}
 \caption{ALMA Observations.}\label{table:1}
 \hline              
 Session name &Configuration &Start time (UTC) & On source & Visibility  & Baseline range & Number of  \\ 
  &  & &   (min.) & calibrator & (m) & antennas \\ 
\endfirsthead
  \hline
\hline 
uid\_\_\_A002\_X97aa1b\_X56e.ms &compact &2014/12/23 22:26:27 & 44.13 & J0239+0416 &  15.05 -- 348.5 & 40\\
uid\_\_\_A002\_X97aa1b\_X95d.ms &compact &2014/12/23 23:49:51 & 44.13 & J0239+0416 &  15.05 -- 348.5 & 40 \\
uid\_\_\_A002\_X85c183\_X1d8c.ms &extended &2014/07/02 09:51:42 &44.87 &	J0301+0118 & 19.6 -- 650.3 & 30\\
uid\_\_\_A002\_X85c183\_X2341.ms &extended &2014/07/02 12:06:23 &41.18  & J0301+0118 &  19.6 -- 650.3 & 30 \\
uid\_\_\_A002\_X87544a\_X27d0.ms &extended &2014/07/21 07:15:31 &44.90 & J0301+0118  &17.8 -- 783.5 & 35 \\
uid\_\_\_A002\_Xa0b40d\_X67c5.ms &extended &2015/05/17 16:38:48 &20.60 & J0239-0234 & 21.4 -- 555.5 & 35\\
\hline
\end{longtable}

\begin{longtable}[htdp]{*{9}{c}}
\caption{Parameters of the identified GMCs}\label{table:2}
%%\hline \\[-5mm] 
%%\hline \\
\hline
ID & dR.A. & dDecl. &  $V_{\rm LSR}$ & $S_{\rm peak}$ & $R$ & $\Delta{v}$ & $M_{\rm vir}$ & $M_{^{13}{\rm CO}}$\\
            &    (arcsec)    &   (arcsec)    & (km s$^{-1}$)  &   (mJy beam$^{-1}$)  &   (pc)  &  (km s$^{-1}$)  & 
($10^5 M_\odot$) & ($10^5 M_\odot$)            \\
(1)      & (2)     & (3)   & (4) & (5)           & (6)            & (7) & (8)        & (9)                        \\ 
\hline 
\endhead
\hline 
\endfoot
\hline 
\multicolumn{9}{l}{\footnotesize (1) cloud ID number (2) \& (3) cloud position from the reference position ((R.A.(J2000) = \timeform{2h42m40.7s},Decl.(J2000),   \timeform{-0D0'47.9''})} \\
\multicolumn{9}{l}{\footnotesize (4) peak velocity (5) peak flux (6) beam-deconvloved cloud radius
(7) velocity width of FWHM of $^{13}$CO} \\ \multicolumn{9}{l}{\footnotesize(8) virial mass, see section 3.4
 (9) LTE mass from $^{13}$CO, see section 3.4} \\

\endlastfoot
% \hline
1& 	-9.1& 	-11.1& 	1007.5& 	37.8& 	211.0& 	19.6& 	152.7& 	420.9 \\ 
2& 	-8.4& 	11.6& 	1078.0& 	33.1& 	195.0& 	11.8& 	50.9& 	183.7 \\ 
3& 	-11.6& 	9.9& 	1052.5& 	31.1& 	191.0& 	21.5& 	164.9& 	293.1 \\ 
4& 	11.9& 	-6.9& 	1235.5& 	23.5& 	189.0& 	17.5& 	108.3& 	162.4 \\ 
5& 	-1.6& 	-13.6& 	1120.0& 	34.1& 	185.0& 	15.0& 	77.9& 	203.9 \\ 
6& 	-3.6& 	-13.4& 	1096.0& 	28.7& 	185.0& 	14.9& 	77.1& 	194.5 \\ 
7& 	-11.4& 	-8.4& 	986.5& 	34.2& 	181.0& 	17.5& 	103.8& 	240.5 \\ 
8& 	15.9& 	-2.1& 	1250.5& 	11.4& 	174.0& 	18.7& 	114.6& 	86.3 \\ 
9& 	19.9& 	10.9& 	1207.0& 	17.3& 	165.0& 	16.5& 	84.9& 	106.2 \\ 
10& 	-4.6& 	-13.1& 	1079.5& 	37.3& 	164.0& 	12.1& 	44.5& 	134.4 \\ 
11& 	7.6& 	-9.9& 	1201.0& 	30.4& 	163.0& 	10.6& 	34.1& 	84.9 \\ 
12& 	-12.1& 	-7.1& 	979.0& 	31.3& 	162.0& 	13.7& 	56.6& 	91.4 \\ 
13& 	-7.1& 	-12.9& 	1040.5& 	37.0& 	162.0& 	15.6& 	73.9& 	183.9 \\ 
14& 	11.9& 	-6.9& 	1232.5& 	23.6& 	161.0& 	9.6& 	27.4& 	58.9 \\ 
15& 	-12.9& 	-5.1& 	964.0& 	29.2& 	156.0& 	16.0& 	75.2& 	97.9 \\ 
16& 	-20.1& 	1.1& 	1027.0& 	16.0& 	156.0& 	24.0& 	168.6& 	91.5 \\ 
17& 	-8.4& 	11.6& 	1081.0& 	32.0& 	156.0& 	6.1& 	10.3& 	39.0 \\ 
18& 	-7.6& 	12.1& 	1088.5& 	33.6& 	154.0& 	16.2& 	76.0& 	112.3 \\ 
19& 	-3.9& 	13.4& 	1111.0& 	23.4& 	154.0& 	13.3& 	50.8& 	81.7 \\ 
20& 	-0.4& 	-12.9& 	1135.0& 	41.9& 	153.0& 	12.1& 	41.4& 	106.6 \\ 
21& 	-5.6& 	-13.6& 	1069.0& 	40.3& 	152.0& 	10.0& 	28.2& 	96.7 \\ 
22& 	2.1& 	14.6& 	1157.5& 	37.6& 	150.0& 	19.4& 	106.2& 	117.1 \\ 
23& 	18.9& 	0.4& 	1249.0& 	18.1& 	149.0& 	14.3& 	57.1& 	53.9 \\ 
24& 	19.1& 	4.1& 	1237.0& 	8.7& 	146.0& 	16.3& 	72.7& 	21.3 \\ 
25& 	15.9& 	6.6& 	1250.5& 	12.6& 	146.0& 	17.3& 	81.4& 	48.5 \\ 
26& 	4.9& 	-11.6& 	1180.0& 	30.1& 	146.0& 	13.5& 	49.6& 	74.4 \\ 
27& 	20.1& 	8.9& 	1223.5& 	16.2& 	145.0& 	9.8& 	25.5& 	46.7 \\ 
28& 	6.9& 	-10.6& 	1189.0& 	33.7& 	145.0& 	13.8& 	51.3& 	79.4 \\ 
29& 	-14.6& 	5.4& 	1028.5& 	19.9& 	144.0& 	9.1& 	22.2& 	43.6 \\ 
30& 	9.4& 	-9.1& 	1214.5& 	25.7& 	144.0& 	14.3& 	54.8& 	67.9 \\ 
31& 	-5.1& 	-18.1& 	1117.0& 	15.5& 	143.0& 	16.8& 	75.2& 	47.3 \\ 
32& 	3.6& 	16.4& 	1139.5& 	18.8& 	142.0& 	15.3& 	62.4& 	58.7 \\ 
33& 	3.1& 	-12.1& 	1165.0& 	28.5& 	142.0& 	10.7& 	30.0& 	54.5 \\ 
34& 	18.6& 	0.1& 	1246.0& 	18.3& 	139.0& 	13.7& 	49.0& 	50.0 \\ 
35& 	6.4& 	13.6& 	1169.5& 	16.5& 	139.0& 	8.5& 	18.3& 	25.9 \\ 
36& 	8.9& 	-9.4& 	1211.5& 	26.5& 	138.0& 	11.5& 	33.9& 	39.7 \\ 
37& 	8.9& 	12.4& 	1196.5& 	24.3& 	138.0& 	6.7& 	11.3& 	26.1 \\ 
38& 	-0.4& 	-12.9& 	1132.0& 	43.0& 	138.0& 	8.0& 	16.1& 	45.1 \\ 
39& 	-9.1& 	-9.6& 	1037.5& 	17.9& 	137.0& 	12.6& 	40.2& 	45.4 \\ 
40& 	-21.1& 	-4.4& 	1004.5& 	13.1& 	137.0& 	9.2& 	21.3& 	31.2 \\ 
41& 	-14.4& 	5.6& 	1042.0& 	16.7& 	136.0& 	10.7& 	28.8& 	34.0 \\ 
42& 	9.6& 	7.4& 	1207.0& 	19.8& 	135.0& 	13.9& 	49.0& 	46.1 \\ 
43& 	-5.1& 	15.4& 	1085.5& 	29.1& 	135.0& 	12.7& 	40.4& 	48.1 \\ 
44& 	2.6& 	-12.4& 	1154.5& 	21.0& 	135.0& 	11.9& 	35.5& 	37.5 \\ 
45& 	-18.4& 	1.4& 	1001.5& 	10.3& 	134.0& 	13.0& 	42.2& 	31.0 \\ 
46& 	-11.9& 	6.6& 	1030.0& 	12.8& 	134.0& 	8.9& 	19.6& 	20.0 \\ 
47& 	-15.1& 	4.9& 	1015.0& 	12.8& 	133.0& 	13.2& 	43.6& 	32.6 \\ 
48& 	-14.4& 	5.6& 	1039.0& 	16.7& 	133.0& 	4.5& 	4.4& 	17.2 \\ 
49& 	-21.9& 	-2.6& 	1013.5& 	11.5& 	133.0& 	11.1& 	30.4& 	32.8 \\ 
50& 	-4.9& 	15.4& 	1091.5& 	28.2& 	133.0& 	7.9& 	15.0& 	33.0 \\ 
51& 	-21.9& 	-8.6& 	1016.5& 	13.4& 	133.0& 	14.3& 	50.4& 	50.2 \\ 
52& 	5.9& 	14.4& 	1162.0& 	19.3& 	131.0& 	8.8& 	18.6& 	25.7 \\ 
53& 	-21.1& 	-4.4& 	1000.0& 	10.7& 	131.0& 	6.8& 	10.7& 	16.1 \\ 
54& 	15.9& 	-4.6& 	1270.0& 	7.7& 	130.0& 	13.4& 	43.3& 	12.7 \\ 
55& 	6.1& 	13.9& 	1151.5& 	12.2& 	129.0& 	6.9& 	11.0& 	10.9 \\ 
56& 	-2.9& 	13.4& 	1121.5& 	26.6& 	127.0& 	6.7& 	10.4& 	21.3 \\ 
57& 	8.6& 	12.6& 	1190.5& 	24.5& 	126.0& 	4.7& 	4.8& 	14.7 \\ 
58& 	14.9& 	4.4& 	1261.0& 	11.8& 	126.0& 	12.6& 	37.3& 	15.1 \\ 
59& 	0.9& 	14.1& 	1133.5& 	15.6& 	123.0& 	13.8& 	44.1& 	33.8 \\ 
60& 	8.9& 	12.1& 	1205.5& 	20.6& 	123.0& 	12.4& 	35.4& 	30.7 \\ 
61& 	-2.1& 	-16.4& 	1129.0& 	8.8& 	122.0& 	11.1& 	27.6& 	8.8 \\ 
62& 	8.6& 	12.6& 	1187.5& 	23.2& 	119.0& 	7.7& 	12.8& 	19.3 \\ 
63& 	12.1& 	8.1& 	1262.5& 	11.6& 	118.0& 	11.2& 	27.5& 	13.8 \\ 
64& 	20.1& 	9.1& 	1216.0& 	17.9& 	118.0& 	9.2& 	18.4& 	21.3 \\ 
65& 	6.4& 	5.6& 	1121.5& 	15.0& 	117.0& 	14.5& 	46.1& 	21.0 \\ 
66& 	10.9& 	9.4& 	1222.0& 	15.5& 	117.0& 	12.0& 	31.3& 	24.9 \\ 
67& 	-22.1& 	-2.9& 	1018.0& 	10.8& 	117.0& 	8.5& 	15.4& 	12.6 \\ 
68& 	-0.6& 	-18.9& 	1142.5& 	12.1& 	116.0& 	11.9& 	30.6& 	20.2 \\ 
69& 	6.6& 	13.1& 	1181.5& 	21.2& 	115.0& 	11.6& 	28.7& 	20.8 \\ 
70& 	-19.4& 	-0.9& 	991.0& 	7.7& 	115.0& 	11.4& 	27.9& 	9.4 \\ 
71& 	-0.9& 	-13.4& 	1159.0& 	13.1& 	114.0& 	12.7& 	34.5& 	14.3 \\ 
72& 	6.9& 	11.1& 	1174.0& 	16.8& 	114.0& 	10.2& 	21.9& 	14.2 \\ 
73& 	11.1& 	3.1& 	1268.5& 	13.6& 	114.0& 	10.6& 	23.7& 	14.6 \\ 
74& 	-10.6& 	-2.6& 	973.0& 	12.9& 	114.0& 	12.6& 	33.5& 	14.7 \\ 
75& 	20.6& 	8.4& 	1240.0& 	9.0& 	113.0& 	9.5& 	18.8& 	6.0 \\ 
76& 	-6.1& 	-11.9& 	1057.0& 	26.0& 	112.0& 	9.8& 	20.0& 	23.3 \\ 
77& 	4.6& 	-18.1& 	1165.0& 	11.1& 	112.0& 	9.0& 	16.8& 	12.5 \\ 
78& 	-14.1& 	-4.9& 	985.0& 	11.0& 	111.0& 	10.1& 	20.7& 	8.1 \\ 
79& 	-6.1& 	-13.9& 	1055.5& 	36.5& 	110.0& 	7.2& 	10.3& 	24.7 \\ 
80& 	-6.4& 	-13.6& 	1051.0& 	36.2& 	110.0& 	7.3& 	10.7& 	22.7 \\ 
81& 	-7.9& 	10.4& 	1049.5& 	8.5& 	109.0& 	11.7& 	27.7& 	5.2 \\ 
82& 	-13.9& 	10.4& 	1025.5& 	7.9& 	109.0& 	11.6& 	27.4& 	5.9 \\ 
83& 	16.4& 	9.1& 	1229.5& 	8.9& 	108.0& 	12.7& 	32.7& 	7.6 \\ 
84& 	-6.4& 	-11.9& 	1054.0& 	26.4& 	107.0& 	9.6& 	18.4& 	20.2 \\ 
85& 	-23.4& 	-7.6& 	1028.5& 	8.2& 	107.0& 	9.2& 	16.6& 	8.4 \\ 
86& 	7.1& 	10.9& 	1162.0& 	8.8& 	107.0& 	8.1& 	12.9& 	4.1 \\ 
87& 	-2.6& 	13.4& 	1127.5& 	23.4& 	106.0& 	5.6& 	5.9& 	5.9 \\ 
88& 	20.4& 	3.9& 	1258.0& 	7.8& 	103.0& 	13.6& 	35.2& 	8.5 \\ 
89& 	-10.4& 	-18.1& 	1076.5& 	23.1& 	102.0& 	16.3& 	50.9& 	29.5 \\ 
90& 	2.9& 	11.6& 	1136.5& 	12.9& 	101.0& 	15.7& 	47.1& 	12.1 \\ 
91& 	-3.6& 	16.1& 	1106.5& 	14.0& 	101.0& 	12.2& 	27.7& 	13.4 \\ 
92& 	11.9& 	13.4& 	1205.5& 	11.5& 	101.0& 	9.6& 	17.0& 	10.7 \\ 
93& 	-2.4& 	13.4& 	1130.5& 	21.0& 	100.0& 	9.9& 	18.2& 	12.8 \\ 
94& 	-13.1& 	-0.1& 	970.0& 	14.8& 	100.0& 	9.2& 	15.4& 	9.3 \\ 
95& 	8.9& 	8.6& 	1172.5& 	16.5& 	99.0& 	11.8& 	25.8& 	13.4 \\ 
96& 	-4.9& 	11.9& 	1081.0& 	10.6& 	98.0& 	14.1& 	36.5& 	8.6 \\ 
97& 	-18.1& 	-0.6& 	979.0& 	11.9& 	98.0& 	8.3& 	12.4& 	7.8 \\ 
98& 	7.4& 	6.9& 	1144.0& 	10.7& 	96.0& 	12.5& 	28.0& 	5.5 \\ 
99& 	-20.9& 	-5.6& 	1022.5& 	8.9& 	95.0& 	9.5& 	16.0& 	6.8 \\ 
100& 	-8.6& 	-10.1& 	1046.5& 	18.1& 	95.0& 	7.8& 	10.5& 	8.6 \\ 
101& 	5.4& 	-9.4& 	1168.0& 	8.5& 	95.0& 	8.6& 	12.9& 	3.0 \\ 
102& 	17.4& 	-1.9& 	1267.0& 	8.5& 	95.0& 	7.9& 	10.8& 	4.2 \\ 
103& 	0.1& 	11.4& 	1102.0& 	13.3& 	95.0& 	5.8& 	5.5& 	5.6 \\ 
104& 	0.1& 	11.4& 	1106.5& 	13.2& 	94.0& 	5.4& 	4.9& 	4.7 \\ 
105& 	-23.1& 	-15.4& 	1037.5& 	8.1& 	94.0& 	10.1& 	17.7& 	6.6 \\ 
106& 	-12.4& 	4.4& 	980.5& 	7.4& 	93.0& 	14.5& 	36.8& 	2.8 \\ 
107& 	-14.9& 	-1.6& 	974.5& 	14.5& 	93.0& 	12.0& 	25.1& 	8.7 \\ 
108& 	-0.9& 	-16.4& 	1136.5& 	7.9& 	93.0& 	9.7& 	16.1& 	3.7 \\ 
109& 	-22.4& 	-3.6& 	1031.5& 	8.6& 	93.0& 	8.3& 	11.7& 	4.6 \\ 
110& 	10.9& 	-3.1& 	1262.5& 	10.8& 	93.0& 	12.0& 	25.0& 	4.2 \\ 
111& 	8.1& 	11.1& 	1166.5& 	10.4& 	93.0& 	8.9& 	13.4& 	4.8 \\ 
112& 	-17.9& 	9.9& 	976.0& 	13.8& 	91.0& 	6.4& 	6.7& 	6.6 \\ 
113& 	7.1& 	7.1& 	1132.0& 	12.8& 	90.0& 	12.6& 	26.5& 	9.1 \\ 
114& 	12.9& 	11.6& 	1225.0& 	11.0& 	90.0& 	9.9& 	16.4& 	6.2 \\ 
115& 	3.6& 	-15.9& 	1156.0& 	10.9& 	90.0& 	9.0& 	13.5& 	4.6 \\ 
116& 	-0.4& 	10.4& 	1093.0& 	7.9& 	90.0& 	8.2& 	11.1& 	3.0 \\ 
117& 	-0.9& 	16.9& 	1114.0& 	10.5& 	90.0& 	7.1& 	8.2& 	3.7 \\ 
118& 	3.4& 	-16.1& 	1153.0& 	11.9& 	90.0& 	8.5& 	11.9& 	4.5 \\ 
119& 	8.6& 	8.4& 	1168.0& 	15.5& 	89.0& 	10.8& 	19.1& 	6.5 \\ 
120& 	-2.9& 	13.1& 	1124.5& 	26.4& 	89.0& 	5.2& 	4.2& 	5.0 \\ 
121& 	-14.6& 	-8.1& 	965.5& 	9.0& 	87.0& 	10.7& 	18.5& 	2.7 \\ 
122& 	6.6& 	11.1& 	1178.5& 	13.6& 	87.0& 	7.3& 	8.5& 	4.2 \\ 
123& 	9.4& 	5.1& 	1223.5& 	8.5& 	86.0& 	10.0& 	16.1& 	2.8 \\ 
124& 	-4.9& 	15.6& 	1088.5& 	28.2& 	86.0& 	5.6& 	4.8& 	6.1 \\ 
125& 	-23.6& 	-10.9& 	1031.5& 	8.3& 	85.0& 	6.0& 	5.4& 	3.8 \\ 
126& 	19.1& 	0.9& 	1273.0& 	8.2& 	83.0& 	12.0& 	22.2& 	4.2 \\ 
127& 	9.4& 	1.6& 	1267.0& 	10.5& 	83.0& 	10.2& 	16.1& 	4.2 \\ 
128& 	6.1& 	-17.9& 	1175.5& 	8.1& 	83.0& 	7.2& 	7.7& 	2.5 \\ 
129& 	10.9& 	8.6& 	1231.0& 	10.4& 	82.0& 	9.3& 	13.1& 	4.2 \\ 
130& 	-8.4& 	-10.1& 	1054.0& 	13.1& 	82.0& 	7.7& 	8.8& 	4.6 \\ 
131& 	-9.1& 	-9.4& 	1042.0& 	18.5& 	82.0& 	7.2& 	7.7& 	4.5 \\ 
132& 	-7.4& 	-7.9& 	1088.5& 	13.1& 	81.0& 	12.0& 	21.6& 	5.4 \\ 
133& 	-17.9& 	10.1& 	968.5& 	8.2& 	81.0& 	3.3& 	1.3& 	1.4 \\ 
134& 	10.6& 	10.4& 	1237.0& 	12.4& 	80.0& 	10.1& 	15.1& 	4.6 \\ 
135& 	12.9& 	-1.1& 	1267.0& 	10.7& 	80.0& 	7.9& 	9.0& 	2.6 \\ 
136& 	-14.6& 	1.9& 	982.0& 	7.9& 	79.0& 	7.9& 	9.0& 	2.7 \\ 
137& 	11.1& 	9.9& 	1250.5& 	18.5& 	77.0& 	7.3& 	7.4& 	4.5 \\ 
138& 	6.4& 	13.9& 	1147.0& 	9.6& 	76.0& 	13.9& 	27.3& 	2.9 \\ 
139& 	11.1& 	9.9& 	1247.5& 	18.2& 	75.0& 	5.2& 	3.5& 	3.1 \\ 
140& 	-17.1& 	9.9& 	982.0& 	12.8& 	75.0& 	7.2& 	6.9& 	2.9 \\ 
141& 	-15.4& 	-9.6& 	980.5& 	10.7& 	74.0& 	11.6& 	18.5& 	3.6 \\ 
142& 	-23.1& 	-15.4& 	1031.5& 	9.5& 	74.0& 	5.4& 	3.7& 	2.6 \\ 
143& 	-16.9& 	9.6& 	985.0& 	13.4& 	73.0& 	10.2& 	13.9& 	4.6 \\ 
144& 	-1.6& 	11.9& 	1088.5& 	10.2& 	73.0& 	8.7& 	10.2& 	2.5 \\ 
145& 	-10.6& 	-14.4& 	1039.0& 	8.4& 	72.0& 	14.4& 	27.9& 	1.9 \\ 
146& 	11.6& 	-1.4& 	1271.5& 	13.4& 	72.0& 	8.3& 	8.9& 	3.2 \\ 
147& 	10.9& 	5.6& 	1253.5& 	8.1& 	69.0& 	16.1& 	33.9& 	1.9 \\ 
148& 	-6.9& 	-5.6& 	1063.0& 	14.5& 	69.0& 	8.4& 	8.9& 	2.8 \\ 
149& 	-18.4& 	1.6& 	994.0& 	7.7& 	68.0& 	4.9& 	2.8& 	1.5 \\ 
150& 	0.9& 	19.1& 	1130.5& 	8.0& 	66.0& 	6.2& 	4.6& 	1.1 \\ 
151& 	8.9& 	-17.6& 	1183.0& 	8.4& 	64.0& 	6.4& 	4.7& 	1.6 \\ 
152& 	-12.1& 	6.1& 	1027.0& 	13.7& 	64.0& 	10.4& 	12.8& 	3.2 \\ 
153& 	-22.6& 	-13.9& 	1028.5& 	8.5& 	63.0& 	9.9& 	11.4& 	3.3 \\ 
154& 	8.9& 	16.6& 	1174.0& 	10.3& 	63.0& 	9.2& 	9.8& 	2.5 \\ 
155& 	-22.9& 	-11.6& 	1021.0& 	7.7& 	60.0& 	10.9& 	13.2& 	1.7 \\ 
156& 	2.6& 	11.4& 	1132.0& 	13.1& 	60.0& 	6.5& 	4.5& 	1.6 \\ 
157& 	7.4& 	-16.4& 	1178.5& 	7.7& 	59.0& 	11.6& 	14.7& 	1.1 \\ 
158& 	-7.4& 	-7.4& 	1079.5& 	10.7& 	59.0& 	7.0& 	5.3& 	1.5 \\ 
159& 	-23.4& 	-10.9& 	1022.5& 	7.7& 	58.0& 	9.1& 	8.9& 	2.4 \\ 
160& 	1.1& 	18.9& 	1135.0& 	10.2& 	58.0& 	6.0& 	3.6& 	1.2 \\ 
161& 	7.6& 	2.9& 	1232.5& 	10.0& 	57.0& 	7.4& 	5.7& 	1.4 \\ 
162& 	6.6& 	5.9& 	1130.5& 	10.4& 	56.0& 	8.2& 	7.0& 	1.4 \\ 
163& 	12.9& 	11.6& 	1232.5& 	11.2& 	56.0& 	4.8& 	2.2& 	1.3 \\ 
164& 	-0.9& 	16.9& 	1109.5& 	12.4& 	54.0& 	2.7& 	0.5& 	0.9 \\ 
165& 	5.1& 	-9.4& 	1165.0& 	9.2& 	54.0& 	4.1& 	1.5& 	0.6 \\ 
166& 	4.9& 	-9.1& 	1160.5& 	9.2& 	53.0& 	5.7& 	3.1& 	0.6 \\ 
167& 	-7.9& 	-9.1& 	1069.0& 	7.7& 	52.0& 	9.4& 	8.4& 	0.9 \\ 
168& 	-6.1& 	-8.9& 	1075.0& 	8.1& 	51.0& 	8.5& 	6.8& 	0.8 \\ 
169& 	9.6& 	5.1& 	1231.0& 	8.9& 	51.0& 	8.5& 	6.7& 	0.7 \\ 
170& 	9.6& 	5.1& 	1228.0& 	9.3& 	49.0& 	2.4& 	0.3& 	0.3 \\ 
171& 	0.9& 	11.6& 	1118.5& 	10.5& 	48.0& 	5.6& 	2.6& 	0.9 \\ 
172& 	-6.6& 	-5.6& 	1066.0& 	13.1& 	48.0& 	4.9& 	2.0& 	0.9 \\ 
173& 	1.1& 	11.6& 	1121.5& 	11.5& 	47.0& 	6.0& 	3.0& 	1.0 \\ 
174& 	0.4& 	5.1& 	1061.5& 	7.5& 	46.0& 	14.0& 	16.7& 	0.7 \\ 
175& 	15.9& 	17.4& 	1165.0& 	8.8& 	45.0& 	6.4& 	3.3& 	0.9 \\ 
176& 	7.4& 	2.6& 	1223.5& 	8.2& 	40.0& 	3.6& 	0.8& 	0.3 \\ 
177& 	-6.9& 	-6.1& 	1070.5& 	10.5& 	39.0& 	4.4& 	1.3& 	0.5 \\ 
178& 	-0.9& 	16.9& 	1105.0& 	10.6& 	38.0& 	4.3& 	1.1& 	0.6 \\ 
179& 	13.1& 	10.9& 	1238.5& 	8.4& 	33.0& 	5.8& 	2.0& 	0.5 \\ 
180& 	-6.9& 	-6.6& 	1075.0& 	10.6& 	29.0& 	6.0& 	1.8& 	0.4 \\ 
181& 	-13.6& 	-15.1& 	1034.5& 	9.9& 	29.0& 	4.2& 	0.9& 	0.5 \\ 
182& 	16.9& 	16.6& 	1177.0& 	7.9& 	27.0& 	9.1& 	4.1& 	0.7 \\ 
183& 	-14.1& 	10.4& 	1019.5& 	8.2& 	25.0& 	11.3& 	5.9& 	0.5 \\ 
184& 	20.1& 	-7.4& 	1243.0& 	6.6& 	25.0& 	5.7& 	1.4& 	0.3 \\ 
185& 	7.4& 	17.6& 	1151.5& 	8.2& 	23.0& 	6.0& 	1.5& 	0.4 \\ 
186& 	11.9& 	-16.9& 	1193.5& 	7.7& 	15.0& 	4.7& 	0.6& 	0.2 \\ 
187& 	-8.9& 	0.6& 	976.0& 	7.0& 	7.0& 	4.8& 	0.3& 	0.2 \\ 

\label{table:GMCcatalogue}
\end{longtable}

\begin{longtable}[htdp]{cccccc}
\caption{Intensities and virial parameter of the identified GMCs}\label{table:3}
%\begin{center}
%\begin{tabular}{ccccccccccccc}
%\hline \\[-5mm] \hline
%\\ 
\hline
ID & $S_{\rm ^{13}CO}$ & $S_{\rm C^{18}O}$ & $S_{\rm CS}$& $S_{\rm CH_3OH}$ &Virial parametrer\\
%[-2mm]
%%ID       & R.A.    & Decl. & $D$ & $V_{\rm LSR}$ & $S_{\rm peak}$ & $R$ & $\sigma_v$ & 
%%$M_{\rm vir} & $M_{\rm 13CO}$ & 13CO(1-0) & C18O(1-0) & CS(2-1) \\[-2mm]
            &     \multicolumn{4}{c}{ (Jy km s$^{-1}$) } &  $M_{^{13}CO}/M_{\rm vir}$  \\
%& (1)      & (2)     & (3)   & (4) & (5)      \\ 
\hline \\%[-6mm]
\endhead
\hline
\endfoot
1& 	3.47& 	1.046& 	0.981& 	0.162& 	2.76 \\ 
2& 	1.746& 	0.529& 	0.372& 	0.229& 	3.61 \\ 
3& 	2.845& 	0.835& 	0.657& 	0.401& 	1.78 \\ 
4& 	1.683& 	0.475& 	0.257& 	0.112& 	1.5 \\ 
5& 	2.181& 	0.692& 	0.478& 	0.166& 	2.62 \\ 
6& 	2.089& 	0.634& 	0.415& 	0.115& 	2.52 \\ 
7& 	2.648& 	0.811& 	0.786& 	0.126& 	2.32 \\ 
8& 	0.978& 	0.244& 	0.158& 	0.056& 	0.75 \\ 
9& 	1.045& 	0.277& 	0.183& 	0.111& 	1.25 \\ 
10& 	1.772& 	0.527& 	0.318& 	0.09& 	3.02 \\ 
11& 	1.183& 	0.368& 	0.163& 	0.072& 	2.49 \\ 
12& 	1.229& 	0.373& 	0.395& 	0.087& 	1.61 \\ 
13& 	2.445& 	0.741& 	0.586& 	0.11& 	2.49 \\ 
14& 	0.814& 	0.238& 	0.119& 	0.053& 	2.15 \\ 
15& 	1.41& 	0.429& 	0.502& 	0.095& 	1.3 \\ 
16& 	1.094& 	0.259& 	0.21& 	0.117& 	0.54 \\ 
17& 	0.558& 	0.17& 	0.121& 	0.055& 	3.8 \\ 
18& 	1.639& 	0.505& 	0.32& 	0.157& 	1.48 \\ 
19& 	1.205& 	0.345& 	0.208& 	0.059& 	1.61 \\ 
20& 	1.635& 	0.523& 	0.308& 	0.133& 	2.58 \\ 
21& 	1.441& 	0.458& 	0.258& 	0.074& 	3.43 \\ 
22& 	1.763& 	0.586& 	0.504& 	0.099& 	1.1 \\ 
23& 	0.734& 	0.179& 	0.173& 	0.139& 	0.94 \\ 
24& 	0.293& 	0.072& 	0.041& 	0.021& 	0.29 \\ 
25& 	0.723& 	0.212& 	0.09& 	0.011& 	0.6 \\ 
26& 	1.253& 	0.4& 	0.168& 	0.059& 	1.5 \\ 
27& 	0.592& 	0.156& 	0.109& 	0.066& 	1.83 \\ 
28& 	1.347& 	0.417& 	0.194& 	0.087& 	1.55 \\ 
29& 	0.687& 	0.178& 	0.12& 	0.072& 	1.97 \\ 
30& 	1.156& 	0.327& 	0.156& 	0.052& 	1.24 \\ 
31& 	0.691& 	0.194& 	0.081& 	0.004& 	0.63 \\ 
32& 	0.917& 	0.257& 	0.139& 	0.033& 	0.94 \\ 
33& 	0.962& 	0.3& 	0.149& 	0.049& 	1.82 \\ 
34& 	0.769& 	0.211& 	0.154& 	0.088& 	1.02 \\ 
35& 	0.443& 	0.134& 	0.108& 	0.02& 	1.42 \\ 
36& 	0.727& 	0.22& 	0.1& 	0.038& 	1.17 \\ 
37& 	0.449& 	0.137& 	0.116& 	0.016& 	2.31 \\ 
38& 	0.831& 	0.277& 	0.162& 	0.096& 	2.8 \\ 
39& 	0.837& 	0.22& 	0.198& 	0.029& 	1.13 \\ 
40& 	0.443& 	0.111& 	0.079& 	0.026& 	1.47 \\ 
41& 	0.599& 	0.163& 	0.111& 	0.031& 	1.18 \\ 
42& 	0.887& 	0.22& 	0.128& 	0.02& 	0.94 \\ 
43& 	0.833& 	0.24& 	0.135& 	0.053& 	1.19 \\ 
44& 	0.715& 	0.217& 	0.121& 	0.044& 	1.06 \\ 
45& 	0.51& 	0.124& 	0.087& 	0.048& 	0.73 \\ 
46& 	0.378& 	0.098& 	0.084& 	0.035& 	1.02 \\ 
47& 	0.581& 	0.139& 	0.093& 	0.043& 	0.75 \\ 
48& 	0.31& 	0.083& 	0.065& 	0.022& 	3.88 \\ 
49& 	0.478& 	0.109& 	0.098& 	0.04& 	1.08 \\ 
50& 	0.588& 	0.176& 	0.101& 	0.025& 	2.21 \\ 
51& 	0.695& 	0.164& 	0.138& 	0.086& 	1.0 \\ 
52& 	0.478& 	0.156& 	0.098& 	0.02& 	1.38 \\ 
53& 	0.246& 	0.053& 	0.044& 	0.025& 	1.5 \\ 
54& 	0.234& 	0.05& 	0.036& 	0.009& 	0.29 \\ 
55& 	0.211& 	0.057& 	0.049& 	0.016& 	1.0 \\ 
56& 	0.437& 	0.136& 	0.067& 	0.007& 	2.05 \\ 
57& 	0.294& 	0.099& 	0.07& 	0.013& 	3.05 \\ 
58& 	0.302& 	0.088& 	0.04& 	-& 	0.4 \\ 
59& 	0.722& 	0.213& 	0.126& 	0.028& 	0.77 \\ 
60& 	0.645& 	0.203& 	0.166& 	-& 	0.87 \\ 
61& 	0.18& 	0.048& 	0.022& 	-& 	0.32 \\ 
62& 	0.425& 	0.135& 	0.11& 	0.015& 	1.51 \\ 
63& 	0.314& 	0.09& 	0.049& 	0.009& 	0.5 \\ 
64& 	0.381& 	0.104& 	0.093& 	0.054& 	1.16 \\ 
65& 	0.543& 	0.115& 	0.082& 	0.031& 	0.45 \\ 
66& 	0.576& 	0.183& 	0.122& 	0.025& 	0.79 \\ 
67& 	0.225& 	0.052& 	0.045& 	0.019& 	0.82 \\ 
68& 	0.415& 	0.12& 	0.027& 	0.012& 	0.66 \\ 
69& 	0.488& 	0.157& 	0.124& 	0.02& 	0.73 \\ 
70& 	0.191& 	0.048& 	0.03& 	0.012& 	0.34 \\ 
71& 	0.351& 	0.103& 	0.054& 	0.012& 	0.41 \\ 
72& 	0.351& 	0.108& 	0.069& 	0.013& 	0.65 \\ 
73& 	0.375& 	0.096& 	0.048& 	0.004& 	0.61 \\ 
74& 	0.385& 	0.095& 	0.058& 	0.006& 	0.44 \\ 
75& 	0.114& 	0.02& 	0.024& 	0.016& 	0.32 \\ 
76& 	0.594& 	0.169& 	0.108& 	0.024& 	1.17 \\ 
77& 	0.274& 	0.075& 	0.024& 	-& 	0.74 \\ 
78& 	0.202& 	0.064& 	0.055& 	0.013& 	0.39 \\ 
79& 	0.616& 	0.197& 	0.134& 	0.038& 	2.4 \\ 
80& 	0.572& 	0.182& 	0.143& 	0.031& 	2.12 \\ 
81& 	0.139& 	0.044& 	0.031& 	0.016& 	0.19 \\ 
82& 	0.14& 	0.029& 	0.031& 	0.016& 	0.22 \\ 
83& 	0.176& 	0.051& 	0.019& 	-& 	0.23 \\ 
84& 	0.546& 	0.153& 	0.106& 	0.018& 	1.1 \\ 
85& 	0.155& 	0.042& 	0.027& 	0.01& 	0.5 \\ 
86& 	0.112& 	0.032& 	0.02& 	0.006& 	0.32 \\ 
87& 	0.162& 	0.049& 	0.022& 	-& 	1.0 \\ 
88& 	0.199& 	0.043& 	0.028& 	0.021& 	0.24 \\ 
89& 	0.686& 	0.215& 	0.105& 	0.03& 	0.58 \\ 
90& 	0.369& 	0.099& 	0.06& 	-& 	0.26 \\ 
91& 	0.365& 	0.099& 	0.049& 	0.016& 	0.48 \\ 
92& 	0.282& 	0.086& 	0.039& 	0.005& 	0.63 \\ 
93& 	0.381& 	0.121& 	0.053& 	0.012& 	0.7 \\ 
94& 	0.282& 	0.09& 	0.038& 	-& 	0.6 \\ 
95& 	0.42& 	0.119& 	0.07& 	0.014& 	0.52 \\ 
96& 	0.27& 	0.077& 	0.053& 	0.011& 	0.24 \\ 
97& 	0.212& 	0.067& 	0.034& 	0.006& 	0.63 \\ 
98& 	0.19& 	0.048& 	0.026& 	0.007& 	0.2 \\ 
99& 	0.17& 	0.037& 	0.02& 	-& 	0.42 \\ 
100& 	0.279& 	0.079& 	0.072& 	0.012& 	0.82 \\ 
101& 	0.103& 	0.023& 	0.016& 	-& 	0.23 \\ 
102& 	0.122& 	0.024& 	0.023& 	0.012& 	0.38 \\ 
103& 	0.192& 	0.054& 	0.028& 	-& 	1.02 \\ 
104& 	0.162& 	0.044& 	0.024& 	-& 	0.98 \\ 
105& 	0.13& 	0.027& 	0.015& 	0.011& 	0.38 \\ 
106& 	0.092& 	0.023& 	0.009& 	-& 	0.08 \\ 
107& 	0.277& 	0.089& 	0.046& 	-& 	0.35 \\ 
108& 	0.116& 	0.026& 	0.015& 	-& 	0.23 \\ 
109& 	0.115& 	0.029& 	0.018&  -& 	0.4 \\ 
110& 	0.149& 	0.036& 	0.013& 	0.005& 	0.17 \\ 
111& 	0.158& 	0.044& 	0.041& 	-& 	0.36 \\ 
112& 	0.186& 	0.049& 	0.03& 	0.013& 	0.99 \\ 
113& 	0.342& 	0.081& 	0.055& 	0.016& 	0.34 \\ 
114& 	0.196& 	0.062& 	0.035& 	-& 	0.38 \\ 
115& 	0.149& 	0.036& 	0.014& 	0.005& 	0.34 \\ 
116& 	0.114& 	0.029& 	0.009& 	-& 	0.28 \\ 
117& 	0.119& 	0.031& 	0.01& 	0.002& 	0.46 \\ 
118& 	0.148& 	0.037& 	0.015& 	-& 	0.38 \\ 
119& 	0.242& 	0.06& 	0.033& 	-& 	0.34 \\ 
120& 	0.179& 	0.052& 	0.029& 	0.005& 	1.18 \\ 
121& 	0.092& 	0.027& 	0.026& 	0.002& 	0.15 \\ 
122& 	0.155& 	0.046& 	0.037& 	0.004& 	0.49 \\ 
123& 	0.111& 	0.018& 	0.009& 	-& 	0.17 \\ 
124& 	0.212& 	0.067& 	0.03& 	0.015& 	1.28 \\ 
125& 	0.093& 	0.019& 	0.012& 	-& 	0.7 \\ 
126& 	0.139& 	0.028& 	0.039& 	0.029& 	0.19 \\ 
127& 	0.181& 	0.033& 	0.032& 	0.006& 	0.26 \\ 
128& 	0.084& 	0.019&  -& 	-& 	0.32 \\ 
129& 	0.167& 	0.047& 	0.029& 	0.005& 	0.32 \\ 
130& 	0.185& 	0.043& 	0.043& 	0.006& 	0.52 \\ 
131& 	0.183& 	0.051& 	0.039& 	-& 	0.59 \\ 
132& 	0.231& 	0.048& 	0.031& 	0.01& 	0.25 \\ 
133& 	0.045& 	0.011& 	0.007& 	0.005& 	1.05 \\ 
134& 	0.181& 	0.06& 	0.05& 	-& 	0.3 \\ 
135& 	0.111& 	0.03& 	0.01& 	-& 	0.29 \\ 
136& 	0.11& 	0.03& 	0.015& -& 	0.3 \\ 
137& 	0.187& 	0.063& 	0.048& 	0.009& 	0.6 \\ 
138& 	0.12& 	0.03& 	0.022& 	0.008& 	0.11 \\ 
139& 	0.135& 	0.046& 	0.038& 	-& 	0.89 \\ 
140& 	0.108& 	0.03& 	0.018& 	0.013& 	0.42 \\ 
141& 	0.144& 	0.047& 	0.024& 	-& 	0.19 \\ 
142& 	0.072& 	0.012& 	0.006& 	-& 	0.72 \\ 
143& 	0.176& 	0.044& 	0.029& 	0.02& 	0.33 \\ 
144& 	0.119& 	0.036& 	0.02& 	-& 	0.24 \\ 
145& 	0.08& 	0.019& 	0.015&  -& 	0.07 \\ 
146& 	0.156& 	0.045& 	0.016& 	-& 	0.35 \\ 
147& 	0.095& 	0.021& 	0.015& 	-& 	0.06 \\ 
148& 	0.152& 	0.027& 	0.014&  -& 	0.32 \\ 
149& 	0.063& 	0.015& 	0.009& 	0.005& 	0.52 \\ 
150& 	0.05& 	0.013& 	0.005& 	0.002& 	0.25 \\ 
151& 	0.07& 	0.015& 	0.005& 	-& 	0.33 \\ 
152& 	0.172& 	0.047& 	0.032& 	0.01& 	0.25 \\ 
153& 	0.114& 	0.02& 	0.014& 	-& 	0.29 \\ 
154& 	0.118& 	0.038& 	0.01& 	-& 	0.26 \\ 
155& 	0.065& 	0.012& 	0.012& 	-& 	0.13 \\ 
156& 	0.096& 	0.03& 	0.014& 	-& 	0.36 \\ 
157& 	0.059& 	0.014& 	0.006& 	-& 	0.08 \\ 
158& 	0.093& 	0.024& 	0.012& 	-& 	0.28 \\ 
159& 	0.092& 	0.017& 	0.016& 	0.008& 	0.27 \\ 
160& 	0.06& 	0.014& 	0.005& 	-& 	0.32 \\ 
161& 	0.096& 	0.022& 	0.012& 	0.002& 	0.25 \\ 
162& 	0.097& 	0.022& 	0.014& 	0.005& 	0.21 \\ 
163& 	0.074& 	0.021& 	0.01& 	-& 	0.6 \\ 
164& 	0.052& 	0.018& 	0.005& 	-& 	1.75 \\ 
165& 	0.039& 	0.009& 	0.005& 	-& 	0.38 \\ 
166& 	0.043& 	0.008& 	0.005& 	-& 	0.2 \\ 
167& 	0.065& 	0.01& 	0.01& 	-& 	0.11 \\ 
168& 	0.055& 	0.011& 	0.008& 	-& 	0.11 \\ 
169& 	0.051& 	0.01& 	0.004& 	-& 	0.11 \\ 
170& 	0.025& 	0.003& 	0.002& 	-& 	1.06 \\ 
171& 	0.066& 	0.019& 	0.007& 	-& 	0.34 \\ 
172& 	0.069& 	0.014& 	0.007& 	-& 	0.45 \\ 
173& 	0.077& 	0.025& 	0.009& 	-& 	0.34 \\ 
174& 	0.057& 	0.012& 	0.009& 	-& 	0.04 \\ 
175& 	0.046& 	0.012& 	0.005& 	0.001& 	0.26 \\ 
176& 	0.028& 	0.005& 	0.003& 	-& 	0.39 \\ 
177& 	0.041& 	0.007& 	0.004& 	-& 	0.36 \\ 
178& 	0.044& 	0.015& 	0.003& 	-& 	0.52 \\ 
179& 	0.037& 	0.012& 	0.006& 	-& 	0.23 \\ 
180& 	0.044& 	0.01& 	0.003& 	-& 	0.23 \\ 
181& 	0.037& 	0.009& 	0.004& 	-& 	0.56 \\ 
182& 	0.046& 	0.013& 	0.005& 	-& 	0.16 \\ 
183& 	0.047& 	0.013& 	0.005& 	-& 	0.09 \\ 
184& 	0.023& 	0.005& 	-& 	-& 	0.21 \\ 
185& 	0.034& 	0.01& 	-& 	-& 	0.27 \\ 
186& 	0.019& 	0.003& 	-& 	-& 	0.37 \\ 
187& 	0.023& 	0.005& 	-& 	-& 	0.69 \\ 
%\end{tabular}
% \end{center}
\end{longtable}

\begin{longtable}{ccccc}
 \caption{parameters of  the identified GMCs}\label{table:4}
\hline 
    & Number &  Average & $\sigma$ & Range  \\
    \endfirsthead
  \hline
\hline 
 $R$ (pc) & 187 & 102 & 40  & 7 - 211 \\
 Velocity width $\Delta v$  (km s$^{-1}$) & & 14.9 & 5.8 & 2.8 - 35.9 \\
$M_{\rm ^{13}CO}$  ($\times 10^6 M_\odot$) & & 2.9 &5.4  & 0.018 - 42 \\
$M_{\rm vir}$  ($\times 10^6 M_\odot$) & &  2.6 &2.9  & 0.027 - 17 \\
Virial parameter $(M_{\rm ^{13}CO}/M_{\rm vir}$) & & 0.85 & 0.80 & 0.04 - 3.9 \\
C$^{18}$O/$^{13}$CO & 187 & 0.27 &0.05  & 0.12 - 0.35 \\
CS/$^{13}$CO &  182 & 0.17 & 0.06  & 0.064 - 0.36 \\
CH$_3$OH/$^{13}$CO& 121 & 0.062 & 0.035  & 0.0007 - 0.21 \\
\hline
\end{longtable}

%% Figure 9

\begin{thebibliography}{}

% Journals(e.g. A\&A,ApJ,AJ,NMRAS,PASP ...)
% Authors, Year, Journal, Vol#, Page#
% Journal Title Abbreviation >> http://www.asj.or.jp/pasj/Jabb.html

\bibitem[Aalto et al.(1995)]{Aalto1995} 
Aalto, S., Booth, R.~S., Black, J.~H., \& Johansson, L.~E.~B.\ 1995, \aap, 300, 369

\bibitem[Aalto et al.(1997)]{Aalto1997} 
Aalto, S., Radford, S.~J.~E., Scoville, N.~Z., \& Sargent, A.~I.\ 1997, \apjl, 475, L107

\bibitem[Aalto et al.(2010)]{Aalto2010} 
Aalto, S., Beswick, R., J\"{u}tte, E.\ 2010, \aap, 522, A59

\bibitem[Aladro et al.(2015)]{Aladro2015} 
Aladro, R., Mart{\'{\i}}n, S., Riquelme, D., et al.\ 2015, \aap, 579, A101 

\bibitem[Alatalo et al.(2016)]{Alatalo2016} 
Alatalo, K., Aladro, R., Nyland, K., et al.\ 2016, \apj, in press (arXiv:1606.07809)


\bibitem[Athanassoula \& Bureau(1999)]{Athanassoula1999} 
Athanassoula, E., \& Bureau, M.\ 1999, \apj, 522, 699 

\bibitem[Baba et al.(2013)]{Baba2013} 
Baba, J., Saitoh, T.~R., \& Wada, K.\ 2013, \apj, 763, 46 

\bibitem[Bachiller et al.(1995)]{Bachiller1995} 
Bachiller, R., Liechti, S., Walmsley, C.~M., \& Colomer, F.\ 1995, \aap, 295, L51 

\bibitem[Bachiller \& P{\'e}rez Guti{\'e}rrez(1997)]{Bachiller1997} 
Bachiller, R., \& P{\'e}rez Guti{\'e}rrez, M.\ 1997, \apjl, 487, L93 

\bibitem[Bland-Hawthorn et al.(1997)]{Bland-Hawthorn1997} 
Bland-Hawthorn, J., Gallimore, J.~F., Tacconi, L.~J., et al.\ 1997, \apss, 248, 9 

\bibitem[Bolatto et al.(2008)]{Bolatto2008} 
Bolatto, A.~D., Leroy, A.~K., Rosolowsky, E., Walter, F., \& Blitz, L.\ 2008, \apj, 686, 948-965 

\bibitem[Bolatto et al.(2013)]{Bolatto2013} 
Bolatto, A.~D., Wolfire, M., \& Leroy, A.~K.\ 2013, \araa, 51, 207


\bibitem[Colombo et al.(2014)]{Colombo2014} 
Colombo, D., Hughes, A., Schinnerer, E., et al.\ 2014, \apj, 784, 3 

\bibitem[Cornwell(2008)]{Cornwell2008} 
Cornwell, T.~J.\ 2008, IEEE Journal of Selected Topics in Signal Processing, 2, 793 

\bibitem[Costagliola et al.(2015)]{Costagliola2015} 
Costagliola, F., Sakamoto, K., Muller, S., et al.\ 2015, \aap, 582, A91 

\bibitem[Danielson et al.(2013)]{Danielson2013} 
Danielson, A.~L.~R., Swinbank, A.~M., Smail, I., et al.\ 2013, \mnras, 436, 2793

\bibitem[Dickman(1978)]{Dickman1978} 
Dickman, R.~L.\ 1978, \apjs, 37, 407 


\bibitem[Donovan Meyer et al.(2012)]{DonovanMeyer2012} 
Donovan Meyer, J., Koda, J., Momose, R., et al.\ 2012, \apj, 744, 42 

\bibitem[Donovan Meyer et al.(2013)]{DonovanMeyer2013} 
Donovan Meyer, J., Koda, J., Momose, R., et al.\ 2013, \apj, 772, 107 

\bibitem[Faesi et al.(2016)]{Faesi2016} 
Faesi, C.~M., Lada, C.~J., \& Forbrich, J.\ 2016, \apj, 821, 125

\bibitem[Flower \& Pineau des For{\^e}ts(2012)]{Flower2012} 
Flower, D.~R., \& Pineau des For{\^e}ts, G.\ 2012, \mnras, 421, 2786 

\bibitem[Fujimoto(1968)]{Fujimoto1968} 
Fujimoto, M.\ 1968, \apj, 152, 391 

\bibitem[Fukui et al.(2008)]{Fukui2008} 
Fukui, Y., Kawamura, A., Minamidani, T., et al.\ 2008, \apjs, 178, 56 

\bibitem[Gallimore et al.(2004)]{Gallimore2004} 
Gallimore, J.~F., Baum, S.~A., \& O'Dea, C.~P.\ 2004, \apj, 613, 794 

\bibitem[Garc{\'{\i}}a-Burillo et al.(2010)]{GarciaBurillo2010} 
Garc{\'{\i}}a-Burillo, S., Usero, A., Fuente, A., et al.\ 2010, \aap, 519, A2 

\bibitem[Garc{\'{\i}}a-Burillo et al.(2014)]{GarciaBurillo2014} 
Garc{\'{\i}}a-Burillo, S., Combes, F., Usero, A., et al.\ 2014, \aap, 567, A125 

\bibitem[Gratier et al.(2012)]{Gratier2012} 
Gratier, P., Braine, J., Rodriguez-Fernandez, N.~J., et al.\ 2012, \aap, 542, A108 

\bibitem[Hatsukade et al.(2015)]{Hatsukade2015} 
Hatsukade, B., Tamura, Y., Iono, D., et al.\ 2015, \pasj, 67, 93 

\bibitem[Helfer \& Blitz(1995)]{Helfer1995} 
Helfer, T.~T., \& Blitz, L.\ 1995, \apj, 450, 90 

\bibitem[Henkel \& Mauersberger(1993)]{Henkel1993} 
Henkel, C., \& Mauersberger, R.\ 1993, \aap, 274, 730 

\bibitem[Heyer et al.(2001)]{Heyer2001} 
Heyer, M.~H., Carpenter, J.~M., \& Snell, R.~L.\ 2001, \apj, 551, 852 

\bibitem[Hirota et al.(2011)]{Hirota2011} 
Hirota, A., Kuno, N., Sato, N., et al.\ 2011, \apj, 737, 40 

\bibitem[Hughes et al.(2010)]{Hughes2010} 
Hughes, A., Wong, T., Ott, J., et al.\ 2010, \mnras, 406, 2065 

\bibitem[Hughes et al.(2013)]{Hughes2013} 
Hughes, A., Meidt, S.~E., Colombo, D., et al.\ 2013, \apj, 779, 46 

\bibitem[Inutsuka et al.(2015)]{Inutsuka2015} 
Inutsuka, S.-i., Inoue, T., Iwasaki, K., \& Hosokawa, T.\ 2015, \aap, 580, A49 

\bibitem[Kennicutt(1998)]{Kennicutt1998} 
Kennicutt, R.~C., Jr.\ 1998, \apj, 498, 541 

\bibitem[Kennicutt \& Evans(2012)]{Kennicutt2012} 
Kennicutt, R.~C., \& Evans, N.~J.\ 2012, \araa, 50, 531

\bibitem[Kikumoto et al.(1998)]{Kikumoto1998} 
Kikumoto, T., Taniguchi, Y., Nakai, N., et al.\ 1998, \pasj, 50, 309 

\bibitem[K{\"o}nig et al.(2016)]{Konig2016} 
K{\"o}nig, S., Aalto, S., Muller, S., et al.\ 2016, \aap, in press (arXiv:1603.05405)

\bibitem[Krips et al.(2011)]{Krips2011} 
Krips, M., Mart{\'{\i}}n, S., Eckart, A., et al.\ 2011, \apj, 736, 37 

\bibitem[Koda et al.(2009)]{Koda2009} 
Koda, J., Scoville, N., Sawada, T., et al.\ 2009, \apjl, 700, L132 

\bibitem[Larson(1981)]{Larson1981} 
Larson, R.~B.\ 1981, \mnras, 194, 809 

\bibitem[Lada \& Lada(2003)]{Lada2003} 
Lada, C.~J., \& Lada, E.~A.\ 2003, \araa, 41, 57 

\bibitem[Leroy et al.(2015)]{Leroy2015} 
Leroy, A.~K., Bolatto, A.~D., Ostriker, E.~C., et al.\ 2015, \apj, 801, 25

\bibitem[Leroy et al.(2016)]{Leroy2016} 
Leroy, A.~K., Hughes, A., Schruba, A., et al.\ 2016, \apj, in press (arXiv:1606.07077)

\bibitem[Lindberg et al.(2016)]{Lindberg2016} 
Lindberg, J.~E., Aalto, S., Muller, S., et al.\ 2016, \aap, 587, A15

\bibitem[Lovas(1992)]{Lovas1992} 
Lovas, F.~J.\ 1992, Journal of Physical and Chemical Reference Data, 21, 181 

\bibitem[MacLaren et al.(1988)]{MacLaren1988} 
MacLaren, I., Richardson, K.~M., \& Wolfendale, A.~W.\ 1988, \apj, 333, 821

\bibitem[Mart{\'{\i}}n et al.(2011)]{Martin2011} 
Mart{\'{\i}}n, S., Krips, M., Mart{\'{\i}}n-Pintado, J., et al.\ 2011, \aap, 527, A36 

\bibitem[Mart{\'{\i}}n et al.(2015)]{Martin2015} 
Mart{\'{\i}}n, S., Kohno, K., Izumi, T., et al.\ 2015, \aap, 573, A116 

\bibitem[Matsushita et al.(1998)]{Matsushita1998} 
Matsushita, S., Kohno, K., Vila-Vilaro, B., Tosaki, T., \& Kawabe, R.\ 1998, \apj, 495, 267

\bibitem[Matsushita et al.(2010)]{Matsushita2010} 
Matsushita, S., Kawabe, R., Kohno, K., Tosaki, T., \& Vila-Vilar{\'o}, B.\ 2010, \pasj, 62, 409

\bibitem[McKee \& Williams(1997)]{Mckee1997}
McKee, C. F. \& Williams, J. P., \apj, 476, 144 

\bibitem[McMullin et al.(2007)]{McMullin2007} 
McMullin, J.~P., Waters, B., Schiebel, D., Young, W., \& Golap, K.\ 2007, Astronomical Data Analysis Software and Systems XVI, 376, 127 

\bibitem[Meier et al.(2000)]{Meier2000} 
Meier, D.~S., Turner, J.~L., \& Hurt, R.~L.\ 2000, \apj, 531, 200 

\bibitem[Meier \& Turner(2004)]{Meier2004} 
Meier, D.~S., \& Turner, J.~L.\ 2004, \aj, 127, 2069

\bibitem[Meier \& Turner(2005)]{Meier2005} 
Meier, D.~S., \& Turner, J.~L.\ 2005, \apj, 618, 259 

\bibitem[Meier \& Turner(2012)]{Meier2012} 
Meier, D.~S., \& Turner, J.~L.\ 2012, \apj, 755, 104 

\bibitem[Meier et al.(2015)]{Meier2015} 
Meier, D.~S., Walter, F., Bolatto, A.~D., et al.\ 2015, \apj, 801, 63 

\bibitem[Muraoka et al.(2009)]{Muraoka2009} 
Muraoka, K., Kohno, K., Tosaki, T., et al.\ 2009, \apj, 706, 1213 

\bibitem[Nakajima et al.(2015)]{Nakajima2015} 
Nakajima, T., Takano, S., Kohno, K., et al.\ 2015, \pasj, 67, 8 

\bibitem[Nishimura et al.(2016)]{Nishimura2016} 
Nishimura, Y., Shimonishi, T., Watanabe, Y., et al.\ 2016, \apj, 818, 161 

\bibitem[Nomura \& Millar(2004)]{Nomura2004} 
Nomura, H., \& Millar, T.~J.\ 2004, \aap, 414, 409 

\bibitem[Ohashi et al.(2014)]{Ohashi2014} 
Ohashi, S., Tatematsu, K., Choi, M., et al.\ 2014, \pasj, 66, 119 

\bibitem[Onodera et al.(2010)]{Onodera2010} 
Onodera, S., Kuno, N., Tosaki, T., et al.\ 2010, \apjl, 722, L127 

\bibitem[Onodera et al.(2012)]{Onodera2012} 
Onodera, S., Kuno, N., Tosaki, T., et al.\ 2012, \pasj, 64, 133

\bibitem[Pan et al.(2015)]{Pan2015} 
Pan, H.-A., Kuno, N., Koda, J., et al.\ 2015, \apj, 815, 59 

\bibitem[Papadopoulos et al.(1996)]{Papadopoulos1996} 
Papadopoulos, P.~P., Seaquist, E.~R., \& Scoville, N.~Z.\ 1996, \apj, 465, 173 

\bibitem[Parsons et al.(2012)]{Parsons2012} 
Parsons, H., Thompson, M.~A., Clark, J.~S., \& Chrysostomou, A.\ 2012, \mnras, 424, 1658 

\bibitem[Petry \& CASA Development Team(2012)]{Petry2012} 
Petry, D., \& CASA Development Team 2012, Astronomical Data Analysis Software and Systems XXI, 461, 849 

\bibitem[Prantzos et al.(1996)]{Prantzos1996} 
Prantzos, N., Aubert, O., \& Audouze, J.\ 1996, \aap, 309, 760

\bibitem[Rand \& Kulkarni(1990)]{Rand1990} 
Rand, R.~J., \& Kulkarni, S.~R.\ 1990, \apjl, 349, L43 

\bibitem[Rebolledo et al.(2012)]{Rebolledo2012} 
Rebolledo, D., Wong, T., Leroy, A., Koda, J., \& Donovan Meyer, J.\ 2012, \apj, 757, 155 

\bibitem[Rebolledo et al.(2015)]{Rebolledo2015} 
Rebolledo, D., Wong, T., Xue, R., et al.\ 2015, \apj, 808, 99 

\bibitem[Rich et al.(2008)]{Rich2008} 
Rich, J.~W., de Blok, W.~J.~G., Cornwell, T.~J., et al.\ 2008, \aj, 136, 2897-2920 

\bibitem[Roberts(1969)]{Roberts1969} 
Roberts, W.~W.\ 1969, \apj, 158, 123 

\bibitem[Rosolowsky \& Blitz(2005)]{Rosolowsky2005a} 
Rosolowsky, E., \& Blitz, L.\ 2005, \apj, 623, 826 

\bibitem[Rosolowsky(2005)]{Rosolowsky2005b} 
Rosolowsky, E.\ 2005, \pasp, 117, 1403 

\bibitem[Rosolowsky \& Leroy(2006)]{RosolowskyLeroy2006} 
Rosolowsky, E., \& Leroy, A.\ 2006, \pasp, 118, 590

\bibitem[Rosolowsky et al.(2008)]{Rosolowsky2008} 
Rosolowsky, E.~W., Pineda, J.~E., Kauffmann, J., \& Goodman, A.~A.\ 2008, \apj, 679, 1338-1351 

\bibitem[Saito et al.(2015)]{Saito2015} 
Saito, T., Iono, D., Yun, M.~S., et al.\ 2015, \apj, 803, 60 

\bibitem[Saito et al.(2016)]{Saito2016} 
Saito, T., Iono, D., Espada, D., et al.\ 2016, \apj, in press (arXiv:1611.01156) 

\bibitem[Sakai et al.(2007)]{Sakai2007} 
Sakai, T., Oka, T., \& Yamamoto, S.\ 2007, \apj, 662, 1043 

\bibitem[Sakamoto et al.(2013)]{Sakamoto2013} 
Sakamoto, K., Aalto, S., Costagliola, F., et al.\ 2013, \apj, 764, 42

\bibitem[Sakamoto et al.(2014)]{Sakamoto2014} 
Sakamoto, K., Aalto, S., Combes, F., Evans, A., \& Peck, A.\ 2014, \apj, 797, 90 

\bibitem[Sanders et al.(1985)]{Sanders1985} 
Sanders, D.~B., Scoville, N.~Z., \& Solomon, P.~M.\ 1985, \apj, 289, 373 

\bibitem[Sawada et al.(2012)]{Sawada2012} 
Sawada, T., Hasegawa, T., \& Koda, J.\ 2012, \apjl, 759, L26

\bibitem[Schmidt(1959)]{Schmidt1959} 
Schmidt, M.\ 1959, \apj, 129, 243 

\bibitem[Scoville \& Sanders(1987)]{Scoville1987} 
Scoville, N.~Z., \& Sanders, D.~B.\ 1987, Interstellar Processes, 134, 21 

\bibitem[Scoville et al.(1988)]{Scoville1988} 
Scoville, N.~Z., Matthews, K., Carico, D.~P., \& Sanders, D.~B.\ 1988, \apjl, 327, L61

\bibitem[Scoville(2013)]{Scoville2013} 
Scoville, N.~Z.\ 2013, Secular Evolution of Galaxies, 491 

\bibitem[Sheth et al.(2008)]{Sheth2008} 
Sheth, K., Vogel, S.~N., Wilson, C.~D., \& Dame, T.~M.\ 2008, \apj, 675, 330-339 

\bibitem[Schinnerer et al.(2000)]{Schinnerer2000} 
Schinnerer, E., Eckart, A., Tacconi, L.~J., Genzel, R., \& Downes, D.\ 2000, \apj, 533, 850 

\bibitem[Schinnerer et al.(2010)]{Schinnerer2010} 
Schinnerer, E., Wei{\ss}, A., Aalto, S., \& Scoville, N.~Z.\ 2010, \apj, 719, 1588 

\bibitem[Schinnerer et al.(2013)]{Schinnerer2013} 
Schinnerer, E., Meidt, S.~E., Pety, J., et al.\ 2013, \apj, 779, 42 

\bibitem[Shimajiri et al.(2014)]{Shimajiri2014} 
Shimajiri, Y., Kitamura, Y., Saito, M., et al.\ 2014, \aap, 564, A68

\bibitem[Solomon et al.(1987)]{Solomon1987} 
Solomon, P.~M., Rivolo, A.~R., Barrett, J., \& Yahil, A.\ 1987, \apj, 319, 730 

\bibitem[Swinbank et al.(2015)]{Swinbank2015} 
Swinbank, A.~M., Dye, S., Nightingale, J.~W., et al.\ 2015, \apjl, 806, L17 

\bibitem[Takano et al.(2014)]{Takano2014} 
Takano, S., Nakajima, T., Kohno, K., et al.\ 2014, \pasj, 66, 75 

\bibitem[Tan et al.(2011)]{Tan2011} 
Tan, Q.-H., Gao, Y., Zhang, Z.-Y., \& Xia, X.-Y.\ 2011, 
Research in Astronomy and Astrophysics, 11, 787

\bibitem[Tosaki et al.(2002)]{Tosaki2002} 
Tosaki, T., Hasegawa, T., Shioya, Y., Kuno, N., \& Matsushita, S.\ 2002, \pasj, 54, 209 

\bibitem[Tsai et al.(2012)]{Tsai2012} 
Tsai, M., Hwang, C.-Y., Matsushita, S., Baker, A.~J., \& Espada, D.\ 2012, \apj, 746, 129 

\bibitem[Tunnard et al.(2015)]{Tunnard2015} 
Tunnard, R., Greve, T.~R., Garcia-Burillo, S., et al.\ 2015, \apj, 800, 25 

\bibitem[Tully(1988)]{Tully1988} 
Tully, R.~B.\ 1988, Cambridge and New York, Cambridge University Press, 1988, 221 p. 

\bibitem[Ueda et al.(2016)]{Ueda2016} 
Ueda, J., Watanabe, Y., Iono, D., et al.\ 2016, \pasj, in press (arXiv:1611.00002) 

\bibitem[Utomo et al.(2015)]{Utomo2015} Utomo, D., Blitz, L., Davis, T., et al.\ 2015, \apj, 803, 16 

\bibitem[van Dishoeck \& Black(1988)]{vanDishoeck1988} 
van Dishoeck, E.~F., \& Black, J.~H.\ 1988, \apj, 334, 771

\bibitem[Vila-Vilaro et al.(2015)]{Vila-Vilaro2015} 
Vila-Vilaro, B., Cepa, J., \& Zabludoff, A.\ 2015, \apjs, 218, 28 

\bibitem[Viti et al.(2011)]{Viti2011} 
Viti, S., Jimenez-Serra, I., Yates, J.~A., et al.\ 2011, \apjl, 740, L3 

\bibitem[Viti et al.(2014)]{Viti2014} 
Viti, S., Garc{\'{\i}}a-Burillo, S., Fuente, A., et al.\ 2014, \aap, 570, A28 

\bibitem[Watanabe et al.(2003)]{Watanabe2003} 
Watanabe, N., Shiraki, T., \& Kouchi, A.\ 2003, \apjl, 588, L121 

\bibitem[Watanabe et al.(2016)]{Watanabe2016} 
Watanabe, Y., Sakai, N., Sorai, K., Ueda, J., \& Yamamoto, S.\ 2016, \apj, 819, 144 

\bibitem[Williams et al.(1994)]{Williams1994} 
Williams, J.~P., de Geus, E.~J., \& Blitz, L.\ 1994, \apj, 428, 693 

\bibitem[Williams \& McKee(1997)]{Williams1997} 
Williams, J.~P., \& McKee, C.~F.\ 1997, \apj, 476, 166 

\bibitem[Wilson \& Matteucci(1992)]{Wilson1992} 
Wilson, T.~L., \& Matteucci, F.\ 1992, \aapr, 4, 1

\bibitem[Wong et al.(2011)]{Wong2011} 
Wong, T., Hughes, A., Ott, J., et al.\ 2011, \apjs, 197, 16 

\bibitem[Yoshida et al.(2015)]{Yoshida2015} 
Yoshida, K., Sakai, N., Tokudome, T., et al.\ 2015, \apj, 807, 66 



\end{thebibliography}
\end{document}